\pdfoutput=1
\documentclass[useAMS,usenatbib]{mn2e}

\usepackage[T1]{fontenc}
\usepackage{aecompl} 
\usepackage{amssymb,amsmath,graphicx}
\usepackage[normalem]{ulem}
\usepackage[usenames]{color}
\newcommand{\avg}[1]{\langle#1\rangle}
\newcommand{\cm}{\ensuremath{\mathrm{cm}}}   
\newcommand{\cmcube}{\ensuremath{\mathrm{cm^{-3}}}} 
\newcommand{\degree}{\ensuremath{^{\circ}}}
\newcommand{\dif}{\mathrm{d}} 
\newcommand{\Av}{\ensuremath{A_\mathrm{v}}} 
\newcommand{\HI}{\ensuremath{\mathrm{H\,\scriptstyle I}}} 
\newcommand{\HII}{\ensuremath{\mathrm{H\,\scriptstyle II}}} 
\newcommand{\HH}{\ensuremath{\mathrm{H}_2}} 
\newcommand{\Ha}{\ensuremath{\mathrm{H}\alpha}} 
\newcommand{\K}{\ensuremath{\mathrm{K}}} 
\newcommand{\kms}{\ensuremath{\mathrm{km}\,\mathrm{s}^{-1}}}
\newcommand{\kpc}{\ensuremath{\mathrm{kpc}}}
\newcommand{\Leff}{\ensuremath{L_\mathrm{eff}}}

\newcommand{\OVI}{\ensuremath{\mathrm{O\,\scriptstyle VI}}} 
\newcommand{\um}{\ensuremath{\mu\mathrm{m}}}
\newcommand{\ngas}{\ensuremath{n}}
\newcommand{\nHI}{\ensuremath{n_\mathrm{H\,\scriptstyle I}}}
\newcommand{\nHH}{\ensuremath{n_\mathrm{H}2}}
\newcommand{\NHI}{\ensuremath{N(\mathrm{H\,\scriptstyle I})}}
\newcommand{\NHH}{\ensuremath{N(\mathrm{H}_2})}

\newcommand{\pc}{\ensuremath{\mathrm{pc}}}
\newcommand{\turb}{\ensuremath{s_\mathrm{max}}}
\title[Gas Density PDFs for M31 and M51]
{Probability distribution functions of gas in M31 and M51}
\author[E. M. Berkhuijsen and A. Fletcher]{E. M. Berkhuijsen$^1$
	\thanks{E-mail: eberkhuijsen@mpifr-bonn.mpg.de} and 
	A. Fletcher$^2$
	\thanks{E-mail: andrew.fletcher@ncl.ac.uk}\\
$^{1}$Max-Planck-Institut f\"ur Radioastronomie, Auf dem H\"ugel 69,
          53121 Bonn, Germany.\\
$^{2}$School of Mathematics and Statistics, Newcastle University, 
          Newcastle upon Tyne, NE1 7RU, U.K.}
\begin{document}

\date{}

\pagerange{\pageref{firstpage}--\pageref{lastpage}} \pubyear{}

\maketitle

\label{firstpage}

\begin{abstract}

We present probability distribution functions (PDFs) of the surface densities
of ionized and neutral gas in the nearby spiral galaxies M31 and M51, as 
well as of dust emission and extinction $A_V$ in M31. The PDFs are close to 
lognormal and those for \HI\ and $A_V$ in M31 are nearly identical. However, 
the PDFs for \HH\ are wider than the \HI\ PDFs and the M51 PDFs have larger 
dispersions than those for M31. We use a simple model to 
determine how the PDFs are changed by variations in the line-of-sight (LOS) 
pathlength $L$ through the gas, telescope resolution and the volume filling 
factor of the gas, $f_v$. In each of these cases the dispersion $\sigma$ of 
the lognormal PDF depends on the variable with a negative power law. We also 
derive PDFs of mean LOS volume densities of gas components in M31 and M51. 
Combining these with the volume density PDFs for different components of the 
ISM in the Milky Way (MW), we find that $\sigma$ decreases with increasing 
length $L$ with an exponent of $-0.76\pm 0.06$, which is steeper than expected.
We show that the difference is due to variations in $f_v$. As $f_v$ is similar 
in M31, M51 and the MW, the density structure in the gas in these galaxies must
be similar. Finally, we demonstrate that an increase in $f_v$ with increasing 
distance to the Galactic plane explains the decrease in $\sigma$ with latitude 
of the PDFs of emission measure and FUV emission observed for the MW.

\end{abstract}

\begin{keywords}
ISM: structure -- turbulence, Galaxies: individual -- M31, M51, MW
\end{keywords}

%________________________________SECTION  1________________________________
\section{Introduction}
\label{sec:introduction}

Many processes can leave an imprint on the complicated structure of the 
interstellar medium (ISM) observed in the Milky Way and nearby galaxies. For 
example, variations in the gas density can be due to: an evolving population 
of gravitationally bound clouds; expanding supernova remnant shells 
compressing the gas; cooling and heating of the gas leading to phase 
transitions; the turbulent and compressible nature of the flow.

During the last two decades compressible turbulence has been
recognized as an important cause of the density structure of the
ISM in the Milky Way. Both isothermal
\citep[][and references therein]{Elmegreen:2004} and multi-phase
\citep{Avillez:2005, Wada:2007} simulations of the ISM produced
lognormal volume density distributions. Several authors noted that for
a large enough volume and for a long enough simulation run, the
physical processes causing the density fluctuations in the ISM in a
galactic disc can be regarded as random, independent, multiplicative
events \citep[e.g.][]{Vazquez-Semadeni:1994, Passot:1998, Wada:2007}. 
Therefore, the PDF of log(density) becomes Gaussian and the density PDF lognormal, 
as long as the local Mach number and the density are not correlated 
\citep{Kritsuk:2007, Federrath:2010}. However,
\citet{Tassis:2010} show that lognormal column density distributions
in molecular clouds are a generic statistical representation of the
inhomogeneous distribution of gas density, whether the inhomogeneity
is a result of turbulence, gravity or ambipolar diffusion. Recent reviews which
discuss density PDFs of molecular clouds and their connection to turbulence and star formation 
include \citet{McKee:2007, Hennebelle:2012, Padoan:2013}
and recent theoretical modelling can be found in \citet*{Federrath:2008, Federrath:2013a, Federrath:2013b}. 

Observational evidence to test the simulation results is slowly
growing. \citet{Padoan:1997}, \citet{Goodman:2009}, \citet{Kainulainen:2009}, 
\citet{Schmalzl:2010}, \citet{Schneider:2012} and 
\citet{Kainulainen:2014} obtained lognormal column density PDFs for
the extinction through dust in molecular clouds in the Milky Way (MW).
The latter authors discuss the relationship between the density PDFs and
star formation. \citet{Schneider:2013} presented lognormal PDFs of 12CO(1-0) 
integrated intensities for four dust clouds in the MW, and \citet{Hughes:2013} 
found lognormal PDFs of the 12CO(1-0) brightness temperature and integrated 
intensities in M51. Also the PDF of the \HH\ column densities
in M33 is lognormal, but with a high-density tail \citep{Druard:2014}.
\citet{Wada:2000} showed that the luminosity function 
of the \HI\ column density in the Large Magellanic Cloud is lognormal.
\citet{Hill:2007, Hill:2008} found lognormal distributions of the
emission measures, which are proportional to the integral of the electron 
density squared, in the Wisconsin \Ha\ Mapper survey
\citep{Haffner:2003} perpendicular to the Galactic plane at latitudes
$|b|>10\degree$, and along the plane, with different parameters. The
emission measures of the extinction-corrected \Ha\ emission from the
galaxy M33 \citep{Tabatabaei:2007} and the volume density of the
dust-bearing gas near stars within $400\,\pc$ from the Sun
\citep{Gaustad:1993} have lognormal distributions. Furthermore, 
\citet{Seon:2011,Seon:2013} obtained lognormal PDFs for the FUV
background emission from the ISRF, and \citet{Bowen:2008} plotted the 
volume density distribution of the $\mathrm{O\,\scriptstyle VI}$-line 
emitting gas in the MW, which is approximately lognormal. Recently, 
\citet{Berkhuijsen:2008, Berkhuijsen:2012} showed that in the Milky Way 
the volume density distribution of the diffuse, ionized gas (DIG) at 
$|b|>5\degree$, and of the diffuse atomic gas are close to lognormal. 
They also discussed differences in the PDF dispersion observed for the 
atomic gas at low and high latitudes, and of warm and cold gas. Here 
we present the PDFs of ionized and neutral gas column densities in the 
nearby galaxies M31 and M51, which are close to lognormal, and discuss 
their properties.

Unfortunately, we cannot separate the variation in gas density due
to turbulence from variations caused by e.g. supernova explosions, HII 
regions or gravitation. Therefore, our PDFs capture all kinds of density 
structure along the lines of sight (LOS), regardless of the origin. 
On the other hand, high-density structures represent only a small 
fraction of the volume of the radio beams at our spatial resolution 
of about 300 \pc. We will interpret the properties of the observed 
PDFs in terms of an effective LOS, $\Leff$, which is the fraction of the 
total LOS occupied by turbulent cells and gas clouds. Furthermore, 
purely for the sake of clarity in the text, we will refer only to clouds 
rather than to clouds or turbulent cells. In reality, we cannot make a firm
distinction. We shall return to this point at the end of the paper.

The paper is organized as follows: The data sets used are described in
Section~\ref{sec:data}, and we show the derived PDFs and discuss their
properties in Section~\ref{sec:PDFs}. Simple models to examine the effects 
of varying resolution, line-of-sight path length and the gas filling factor 
on the form of the PDFs are presented in Section~\ref{sec:models}. We compare 
the observed properties of the PDFs to those expected from simulation studies 
in Section~\ref{sec:discussion}, and in Section~\ref{sec:summary} we 
summarize our results.

%_________________________SECTION  2_______________________________________
\section{The data}
\label{sec:data}

M31, the spiral galaxy nearest to us, is at a distance of $780\pm
40\,\kpc$ \citep{Stanek:1998}. The Sb galaxy of large angular extent
(more than $3\degree \times 1\degree$) and low surface brightness has
an inclination of $i=77.5\degree$ \citep{Braun:1991, Chemin:2009} and
so is nearly edge-on. The position angle is $PA = 37.7\degree$\,.
Our gas density PDFs are based on the Westerbork \HI\
survey of \citet{Brinks:1984} at an angular resolution of
$24\times 36\,\rm{arcsec^2}$, corrected for missing spacings, the IRAM
12CO(1-0) survey of \citet{Nieten:2006} at a resolution of
23 arcsec, and the extinction corrected \Ha\ data of
\citet{Devereux:1994}, at a resolution of a few arcsec, produced by
\citet{Tabatabaei:2010}. \citet{Nieten:2006} converted the CO data to
column densities of molecular gas using the constant conversion factor
$X_\mathrm{CO} = 1.9\times 10^{20}\,\mathrm{mol}\,\cm^{-2}\,(\K\,\kms)^{-1}$.
Before deriving the PDF we smoothed each map to an angular resolution
of 45\,arcsec, corresponding to $170\,\pc \times 785\,\pc$ along
major $\times$ minor axis in the plane of M31. Each PDF was calculated 
from independent data points (at least $1.67\times$ beamwidth apart) 
above $2\times$ rms noise.

M51 is a bright Sbc galaxy at a distance of 7.6\,Mpc
\citep{Ciardullo:2002}. It has an angular extent of about 10\,arcmin,
a position angle of $PA = -10\degree$\, and an inclination 
of $i=20\degree$ \citep{Tully:1974, Pety:2013}. For the PDFs we 
used the \HI\ data from the THINGS survey \citep{Walter:2008} 
at a resolution of $5.82 \times 5.56\,\rm{arcsec^2}$, the BIMA 12CO(1-0)
survey of \citet{Helfer:2003} at a resolution of $5.8 \times
5.1\,\rm{arcsec^2}$, which was corrected for missing spacings using
single-dish observations, and the \Ha\ map of \citet{Greenawalt:1998}
at a resolution of a few arcsec. All maps were reduced to a common
field of about $8.5 \times 10.5\,\rm{arcmin^2}$ covering most of M51
apart from the extended southern outer arm in \HI. They were then
smoothed to a resolution of 8\,arcsec corresponding to $295 \times 315\,\pc^2$
along major $\times$ minor axis in the plane of M51. We converted the 
CO data to column densities of \HH\ using the conversion factor 
$4.75\times 10^{19}\,\mathrm{mol}\,\cm^{-2}\,(\K\,\kms)^{-1}$,
one quarter of that for the Milky Way \citep{Guelin:1995}. Of each map,
only independent data points above $2\times$ rms noise were used to
produce the PDF.

%________________________SECTION  3________________________________________

\section{Probability distribution functions}
\label{sec:PDFs}

In this Section we present the observed surface density PDFs
\footnote{We call our PDFs of the gas densities probability distribution 
functions, rather than probability density functions, to avoid referring to 
the ``probability density of the density''.} 
for M31 and M51 
and characterise them by fitting lognormal distributions to the histograms. 
Since there are various ways to paramaterise a lognormal we shall briefly 
describe the exact equations we use. To allow for easy comparison with 
\citet{Berkhuijsen:2008} and other work on observed PDFs we use common, or 
base 10, logarithms. The lognormal probability distribution function for the 
random variable $x$ is given by
\begin{equation}
\label{eq:lognormal}
\Lambda(x|\mu, \sigma)=\frac{\log_\mathrm{10}e}{\sigma x\sqrt{2\pi}}
\exp{[-(\log_\mathrm{10}{x}-\mu)^2/2\sigma^2]}\,\mathrm{d}x,
\end{equation}
which is equivalent to a Gaussian distribution for $\log_\mathrm{10}(x)$,
\begin{multline}
\label{eq:gaussian}
N[\log_\mathrm{10}(x)|\mu, \sigma] = \\ 
 \frac{1}{\sigma \sqrt{2\pi}}
\exp{[-(\log_\mathrm{10}{x}-\mu)^2/2\sigma^2]}\,\mathrm{d}\log_\mathrm{10}x.
\end{multline}
We use Eq.~\ref{eq:gaussian} to derive $\mu$ and $\sigma$ from fits to 
histograms of $\log_\mathrm{10}(x)$, taking into account the logarithmic 
bin-width to ensure that the integral of the PDF is unity. It is important 
to note that $\mu$ and $\sigma$ are the mean and standard deviation of the 
Gaussian distribution of $\log_\mathrm{10}(x)$; the mean and standard 
deviation of the distribution of $x$ are given by 
\begin{displaymath}
E[x]=\exp{(\mu + \sigma^2/2)},
\end{displaymath} 
and 
\begin{displaymath}
S[x]=\sqrt{(\exp{\sigma^2}-1)\exp{(2\mu+\sigma^2)}},
\end{displaymath}
respectively. Thus the variance of $x$ depends on \emph{both} the mean and 
standard deviation of $\log_\mathrm{10}(x)$. We shall try to avoid confusion 
by calling $\sigma$ the dispersion of the PDF and using $X_{0}=10^{\mu}$ to 
describe the peak of the PDF (note that $10^{\mu}$ is the median of the 
distributions of both $x$ and $\log_\mathrm{10}(x)$). For the Gaussian PDFs 
of $\log_\mathrm{10}(x)$, which is the form we show in our figures, small 
(large) values of $\sigma$ result in a narrow (broad) Gaussian PDF.  

%___________________________________________SECTION  3.1____________________

\subsection{PDFs of surface densities}
\label{subsec:surface}

%_____________FIGURE_1___________
\begin{figure}
\begin{center}
\includegraphics[width=0.45\textwidth]{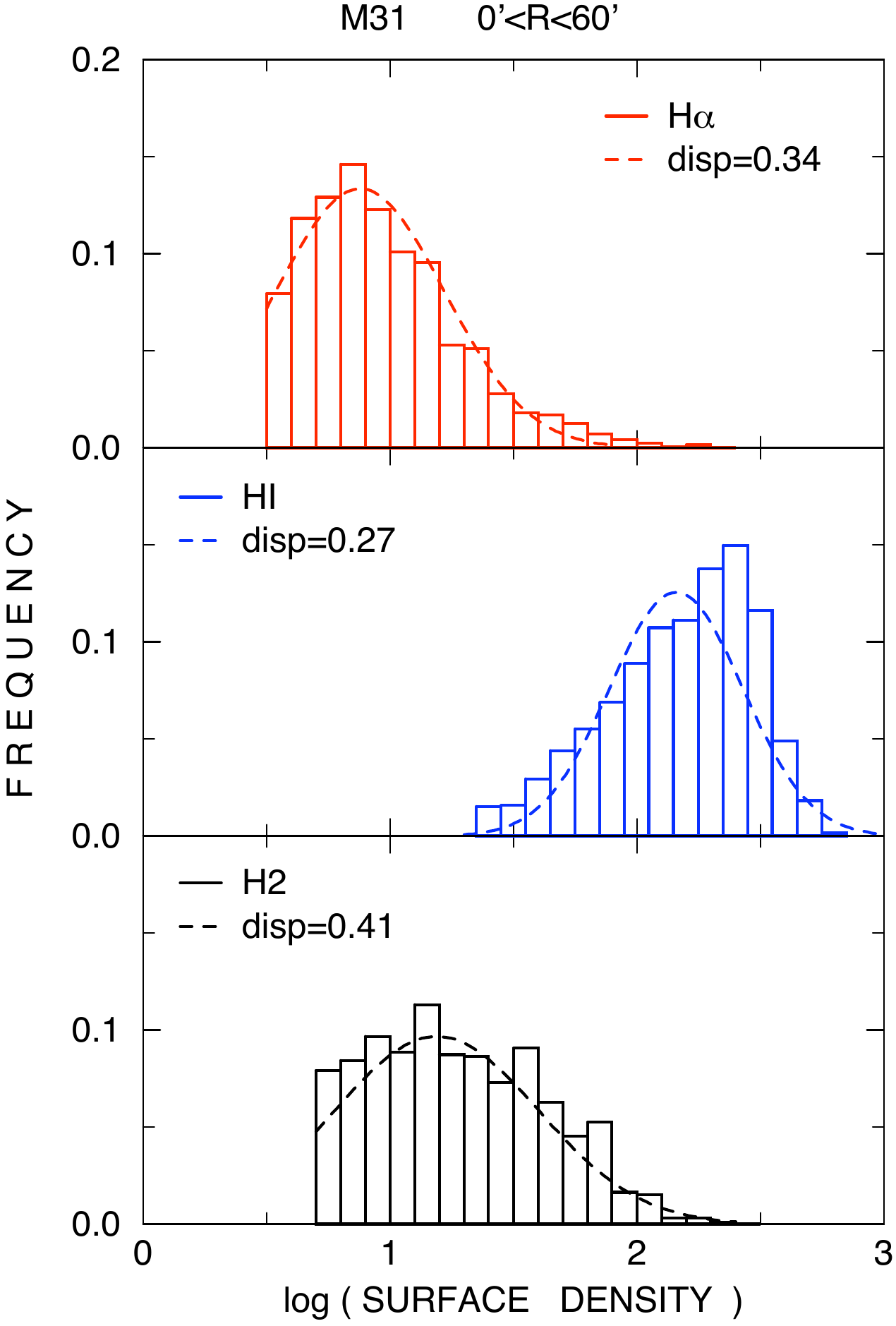}
\end{center}
\caption{
Probability distribution functions of the surface densities of ionized and 
neutral gas in the radial interval $0\arcmin<R<60\arcmin$ in M31. \emph{Top}: 
\Ha\ (in $\cm^{-6}\,\pc$), \emph{middle}: \NHI\ (in $10^{19}\,\mathrm{atom}\,
\cm^{-2}$), \emph{bottom}: \NHH\ (in $10^{19}\,\mathrm{mol}\,\cm^{-2}$). Dashed 
lines show the lognormal fits given in Table~\ref{table:M31:surface}.
}
\label{fig:M31}
\end{figure}

%_____________FIGURE_2___________
\begin{figure}
\begin{center}
\includegraphics[width=0.46\textwidth]{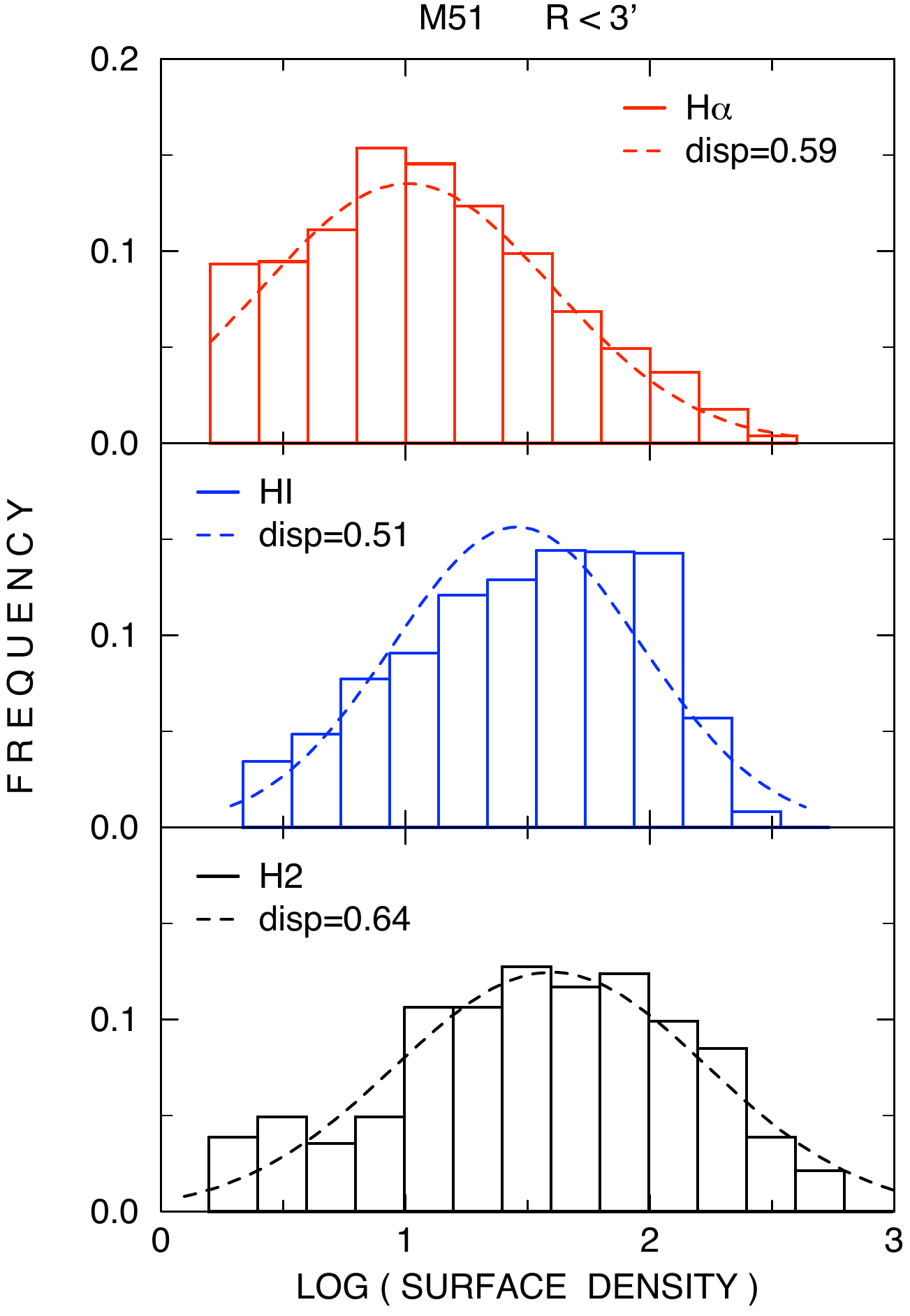}
\end{center}
\caption{Probability distribution functions of the surface densities of 
ionized and neutral gas in M51. \emph{Top}: \Ha\ (in $\cm^{-6}\,\pc$), 
\emph{middle}: \NHI\ (in $10^{19}\,\mathrm{atom}\,\cm^{-2}$), \emph{bottom}: 
\NHH\ (in $10^{19}\,\mathrm{mol}\,\cm^{-2}$). Dashed lines show the lognormal 
fits given in Table~\ref{table:M51}.}
\label{fig:M51}
\end{figure}

In Fig.~\ref{fig:M31} we present the M31 PDFs of the emission measures
of the ionized gas and of the column densities of \HI\ and \HH\ for
the radial range $0\arcmin<R<60\arcmin$ in the plane of the galaxy,
corresponding to $0<R<13.6\,\kpc$. In Table~\ref{table:M31:surface} we
list the parameters of the lognormal fits together with those for the
area containing the bright emission ring ($35\arcmin<R<60\arcmin$ or
$7.9<R<13.6\,\kpc$) and for the weaker emission interior to this 'ring'
($0\arcmin<R<35\arcmin$). For \Ha\ and \HH\ the fits of the PDFs for
the subregions are nearly the same, but for \HI\ the position of the
peak of the PDF, $X_0$, for the bright 'ring' is at a significantly
higher density than for the inner region and the dispersion is
smaller. This difference in $X_0$ causes the low-density excess 
--- and hence the asymmetry --- in the total PDF and its larger 
dispersion compared to that of the subregions.

Figure~\ref{fig:M51} shows the M51 PDFs of the emission measures of
the ionized gas and of the column densities of \HI\ and \HH\ at radii
$R<3\arcmin$. The parameters of the lognormal fits are given in 
Table~\ref{table:M51}. Again the \HI\ PDF has an excess at low densities.
The peak at high densities is due to gas in the outer spiral arms. 
We have tried to construct separate PDFs for spiral arms and the interarm 
regions by using a spiral-arm mask; this was created using 
the coefficients of a wavelet transform of the CO map with a scale of
about $1\,\kpc$, the positive coefficients tracing a clear two-armed
spiral. Although the dispersions of these PDFs could not be reliably
determined, the values of $X_0$ of the arm PDFs are significantly
higher than of the PDFs for the interarm regions. Recently, 
\citet{Hughes:2013} found that PDFs of the integrated 12CO(1-0) intensities
in M51 for arm regions are clearly wider than for interarm regions, and
have an about 25 per cent larger $X_0$. They also presented PDFs for the inner 
$4.5\arcmin \times 2.8\arcmin$ of M51. At a spatial resolution of $53\,\pc$, 
using all pixels with an intensity above 3 times the noise, they obtained a 
dispersion of $\sigma = 0.62$, the same as our value for a larger area of 
M51 observed at a lower resolution.

%_____________TABLE_1__________________________________________________
\begin{table*}
\caption{
Lognormal fits to the PDFs of gas surface densities in M31, HPBW $45\arcsec$. 
}
\begin{tabular}{llrrrrrr}
\hline
 & & & & \multicolumn{2}{c}{Position of maximum} & Dispersion & \\
$X$ & Units & $R(\arcmin)$ & $N$ & $\mu$ & $X_0$ & $\sigma$ & $\chi^2$  \\
\hline
\NHI & $10^{19}\,\mathrm{atom}\,\cm^{-2}$ & 0--60 & 1249 & $2.16\pm0.03$ & 
	$145\pm10$ & $0.27\pm0.02$ & 12.7 \\
 & & 0--35 & 368 & $1.86\pm0.01$ & $72\pm2$ & $0.23\pm0.01$ & 0.8 \\
 & & 35--60 & 881 & $2.32\pm0.02$ & $209\pm10$ & $0.19\pm0.01$ & 5.5 \\
 & & & & & & & \\
\NHH & $10^{19}\,\mathrm{mol}\,\cm^{-2}$ & 0--60 & 971 & $1.19\pm0.04$ & 
	$15.5\pm1.5$ & $0.41\pm0.02$ & 2.9 \\
 & & 0--35 & 250 & $1.11\pm0.05$ & $12.9\pm1.6$ & $0.39\pm0.04$ & 1.5 \\
 & & 35--60 & 721 & $1.22\pm0.04$ & $16.6\pm1.6$ & $0.40\pm0.03$ & 3.0 \\
 & & & & & & & \\
\Ha & $\cm^{-6}\,\pc$ & 0--60 & 1097 & $0.88\pm0.03$ & 
	$7.6\pm0.5$ & $0.34\pm0.02$ & 2.9 \\
 & & 0--35 & 397 & $0.93\pm0.03$ & $8.5\pm0.6$ & $0.34\pm0.02$ & 1.2 \\
 & & 35--60 & 700 & $0.85\pm0.03$ & $7.1\pm0.5$ & $0.34\pm0.02$ & 2.3 \\

\hline\noalign\\  
\end{tabular}

\medskip
N is the total number of data points; the error in each bin $\delta_i$ is 
estimated as $\delta_i=\sqrt{N_i}$ for (number of bins$-2$) degrees of 
freedom; $\chi^2$ is the reduced chi-squared goodness of fit parameter.
\label{table:M31:surface}
\end{table*}

%_____________TABLE_2___________

\begin{table*}
\caption{
Lognormal fits to the PDFs of gas surface densities in M51 at $R<3\arcmin$, HPBW $8\arcsec$.
}
\begin{tabular}{llrrrrr}
\hline
 & & & \multicolumn{2}{c}{Position of maximum} & Dispersion & \\
$X$ & Units & $N$ & $\mu$ & $X_0$ & $\sigma$ & $\chi^2$  \\
\hline
\NHI & $10^{19}\,\mathrm{atom}\,\cm^{-2}$ & 2015 & $1.46\pm0.05$ & 
	$28.7\pm3.3$ & $0.51\pm0.05$ & 6.4 \\
\NHH & $10^{19}\,\mathrm{mol}\,\cm^{-2}$ & 282 & $1.60\pm0.05$ & 
	$39.4\pm4.5$ & $0.64\pm0.05$ & 1.1 \\
\Ha & $\cm^{-6}\,\pc$ & 728 & $1.01\pm0.04$ & 
	$10.2\pm1.0$ & $0.59\pm0.03$ & 1.4 \\
\hline\noalign\\  
\end{tabular}

\medskip
N is the total number of data points; the error in each bin $\delta_i$ 
is estimated as $\delta_i=\sqrt{N_i}$ for (number of bins$-2$) degrees of 
freedom; $\chi^2$ is the reduced chi-squared goodness of fit parameter.
\label{table:M51}
\end{table*}

The PDFs of \Ha, \HI\ and \HH\ for M31 and M51 presented in
Figs.~\ref{fig:M31} and \ref{fig:M51} are close to lognormal. That the
PDFs of \HH\ are lognormal may seem surprising in view of the recent
work of \citet{Shetty:2011}. In their simulations of molecular clouds,
the authors found that the PDFs of 12CO(1-0)-intensities and of CO
column densities are generally not lognormal and have different shapes
when CO is optically thick. But at low densities and for the model of
a Milky Way cloud of medium density the PDFs are lognormal. The box
size in their simulations is $20\,\pc$, much smaller than the typical
resolution of $300\,\pc$ used for the PDFs of the gas surface
densities in M31 and M51. Hence, the line intensity seen by the radio
beam is the average intensity of a large sample of individual clouds.
Saturated lines will hardly be observed because the filling factor of
the densest clouds is very small. Therefore the observed CO PDFs
become lognormal and will give a fair representation of 
the PDF of the \HH\ column density.

Comparing Figures~\ref{fig:M31} and \ref{fig:M51} we note the following:

(a) Both PDFs of \Ha\ show a low-intensity cut-off at the sensitivity limit. 
A similar cut-off occurs in the PDF of \HH\ for M31.

(b) The \Ha\ PDF for M51 extends to higher emission measures than that
for M31 and the emission measure of the maximum is about 35\% higher
than that of M31. Since the M51 data were not corrected for extinction
and the inclination of M51 is smaller than that of M31, this means
that the face-on surface brightness in \Ha\ of M51 is much larger than
that of M31.

(c) The steep decrease seen in the \HI\ PDFs at high column densities
may be attributed to opacity in the \HI\ lines. In scatter plots of
star-formation rate against \HI\ mass surface density a cut-off near
$10\,\mathrm{M}_\odot\,\pc^{-2}$ is visible in M31
\citep{Tabatabaei:2010}, M51 and many other galaxies
\citep{Bigiel:2008}. \citet{Braun:2009} have shown that after
correcting the HI column densities of M31 for opacity the apparent
cut-off vanishes.

(d) The \HI\ column densities of M31 extend to higher values than
those of M51 and the maximum of the PDF of M31 occurs at a column
density that is 5 times that of M51. On the sky, M31 is much brighter
in \HI\ than M51, but the face-on value of $X_0$, defined as $X_{0}\cos{i}$, 
of M31 is only about 1.2 times that of M51. The \HI\ PDF for M51 has a 
pronounced low-density tail, while that for M31 only shows a mild low-density 
excess. The low-density tail of the M51 PDF may be a relic of the last encounter
with the companion galaxy NGC5195 \citep{Howard:1990} that gave rise to the 
extended \HI\ arm in the south-east \citep{Rots:1990}.

(e) The \HH\ column densities of M51 are much higher than of M31: the face-on 
value of $X_0$ of M51 is 11 times that of M31. The star formation rate of 
M51 is five times higher than that of M31 \citep{Tabatabaei:2013}, which may be
related to the higher column densities reached in M51 than in M31 
\citep{Kainulainen:2014}. \HH\ is the dominant gas phase in the inner part of 
M51 \citep{Garcia-Burillo:1993, Schuster:2007}, while in M31 the atomic gas is 
the dominant component \citep{Dame:1993, Nieten:2006}.

%____________________________________________SECTION  3.2____________________

\subsection{PDFs of volume densities in M31}
\label{subsec:volume}

%_____________FIGURE_3___________
\begin{figure}
\begin{center}
\includegraphics[width=0.45\textwidth]{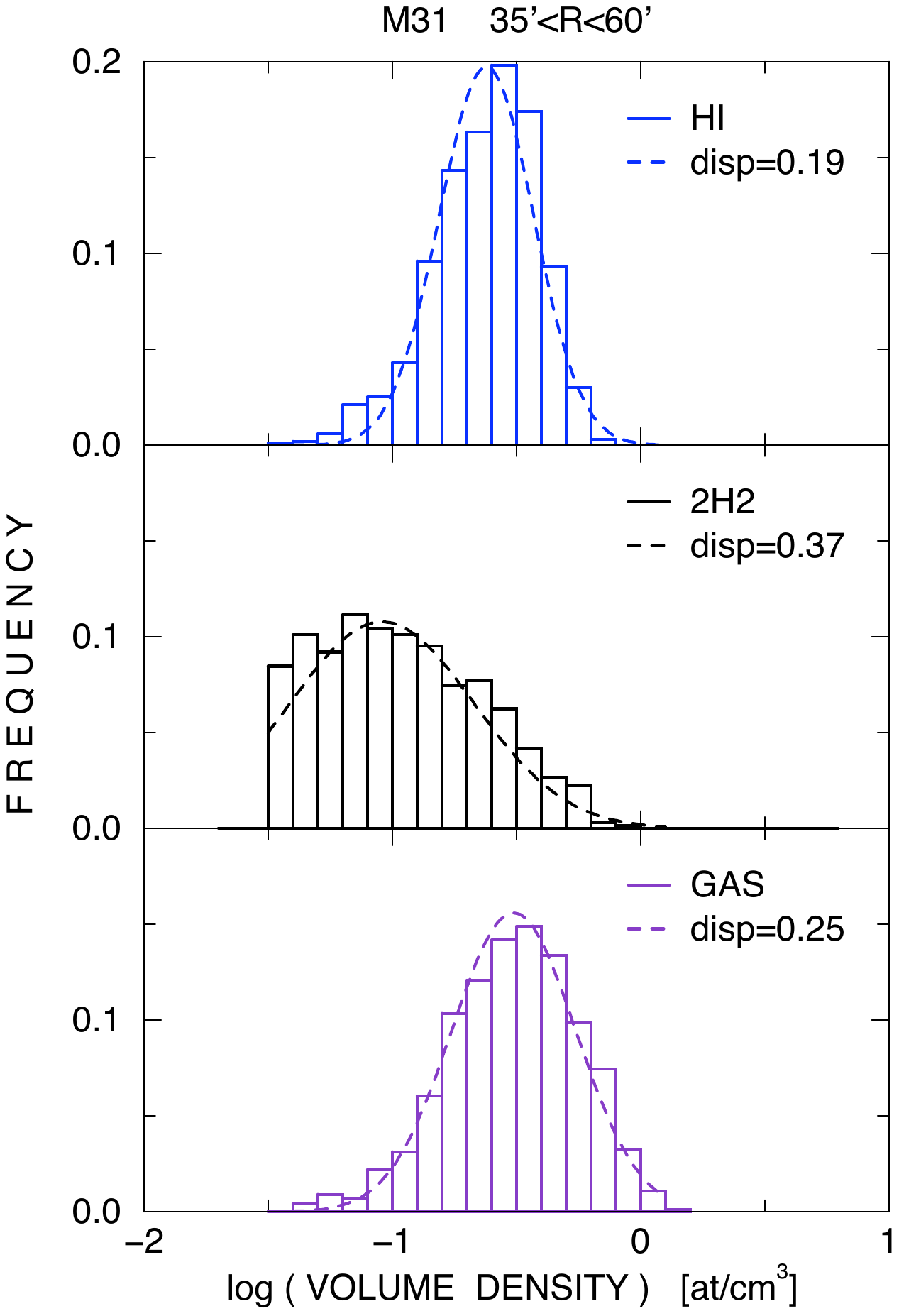}
\end{center}
\caption{Probability distribution functions of mean volume densities along 
the line of sight (in $\cmcube$) of neutral gas in the radial interval 
$35\arcmin<R<60\arcmin$, the bright emission 'ring', in M31. 
\emph{Top}: $\avg{\nHI}$, \emph{middle}: $\avg{2\nHH}$, \emph{bottom}: 
$\avg{\ngas}=\avg{\nHI}+\avg{2\nHH}$. Dashed lines show the lognormal fits 
given in Table~\ref{table:M31:volume}.}
\label{fig:M31:ring}
\end{figure}

%_____________TABLE_3___________

\begin{table*}
\caption{Lognormal fits to the PDFs of gas volume densities in M31, 
HPBW $45\arcsec$. 
}
\begin{tabular}{llrrrrrr}
\hline
 & & & & \multicolumn{2}{c}{Position of maximum} & Dispersion & \\
$X$ & Units & $R(\arcmin)$ & $N$ & $\mu$ & $X_0$ & $\sigma$ & $\chi^2$  \\
\hline
$\avg{\nHI}$ & $\mathrm{atom}\,\cm^{-3}$ & 0--60 & 1457 & $-0.75\pm0.03$ & 
	$0.178\pm0.012$ & $0.25\pm0.02$ & 12.5 \\
 & & 0--35 & 459 & $-1.00\pm0.02$ & $0.100\pm0.005$ & $0.23\pm0.01$ & 1.6 \\
 & & 35--60 & 998 & $-0.62\pm0.02$ & $0.240\pm0.011$ & $0.19\pm0.01$ & 5.7 \\
 & & & & & & & \\
$\avg{n_{2\HH}}$ & $\mathrm{atom}\,\cm^{-3}$ & 0--60 & 920 & $-1.05\pm0.03$ & 
	$0.089\pm0.006$ & $0.39\pm0.02$ & 2.6 \\
 & & 0--35 & 248 & $-1.08\pm0.05$ & $0.083\pm0.010$ & $0.41\pm0.03$ & 1.3 \\
 & & 35--60 & 672 & $-1.04\pm0.03$ & $0.091\pm0.007$ & $0.37\pm0.02$ & 2.0 \\
 & & & & & & & \\
$\avg{\ngas}$ & $\mathrm{atom}\,\cm^{-3}$ & 0--60 & 1451 & $-0.61\pm0.02$ & 
	$0.245\pm0.012$ & $0.30\pm0.01$ & 5.1 \\
 & & 0--35 & 457 & $-0.82\pm0.02$ & $0.151\pm0.007$ & $0.32\pm0.01$ & 1.1 \\
 & & 35--60 & 994 & $-0.51\pm0.01$ & $0.309\pm0.007$ & $0.25\pm0.01$ & 2.6 \\

\hline\noalign\\  
\end{tabular}

\medskip
$\chi^2$ is the reduced chi-squared goodness of fit parameter: the error in 
each bin $\delta_i$ is estimated as $\delta_i=\sqrt{N_i}$ for (number of 
bins$-2$) degrees of freedom.
\label{table:M31:volume}
\end{table*}

We calculated the mean \HI\ volume densities, $\avg{\nHI}$, from the column 
densities using the formula for the \HI\ scale height increasing with radius, 
$h(R)$, given by \citet{Braun:1991}, scaled to the distance $D=780\,\kpc$. 
This gives $h(10\arcmin)=240\,\pc$ and $h(60\arcmin)=425\,\pc$. The volume 
density then is $\avg{\nHI}=\NHI/L$ where the line of sight $L=2h(R)/\cos{i}$. 
Assuming a scale height of the molecular gas equal to $h(R)/2$ yielded the 
mean volume densities of the molecular gas, $\avg{n_{2\HH}}$. The volume 
density of all neutral gas then is $\avg{\ngas} = \avg{\nHI} + \avg{n_{2\HH}}$. 
As an example, we present the PDFs of these volume densities 
for the 'ring' in Fig.~\ref{fig:M31:ring} and the lognormal fits for all 
regions in Table~\ref{table:M31:volume}. The \HI\ PDF is narrower and more 
symmetric than that for the full area shown in Fig.~\ref{fig:M31} because the 
low-density extension is not included. It is shifted by a factor of 2.4 w.r.t.
the PDF for the inner region. Comparison with Table~\ref{table:M31:surface} 
shows that the dispersions of the PDFs of $\avg{\nHI}$ and $\avg{n_{2\HH}}$ 
are the same as those of the column density PDFs within the errors (see 
Section~\ref{sec:models} for a simple explanation for this equivalence.). 
These estimates for the mean volume density PDFs are useful in that they 
allow us to compare the properties of M31 with those of the Milky Way in 
Section~\ref{subsec:LOS}.

%__________________________________SECTION  4______________________________

\section{Dependence of density PDF parameters on pathlength, resolution and 
filling factor}
\label{sec:models}

%_____________FIGURE_4___________
\begin{figure}
\begin{center}
\includegraphics[width=0.4\textwidth]{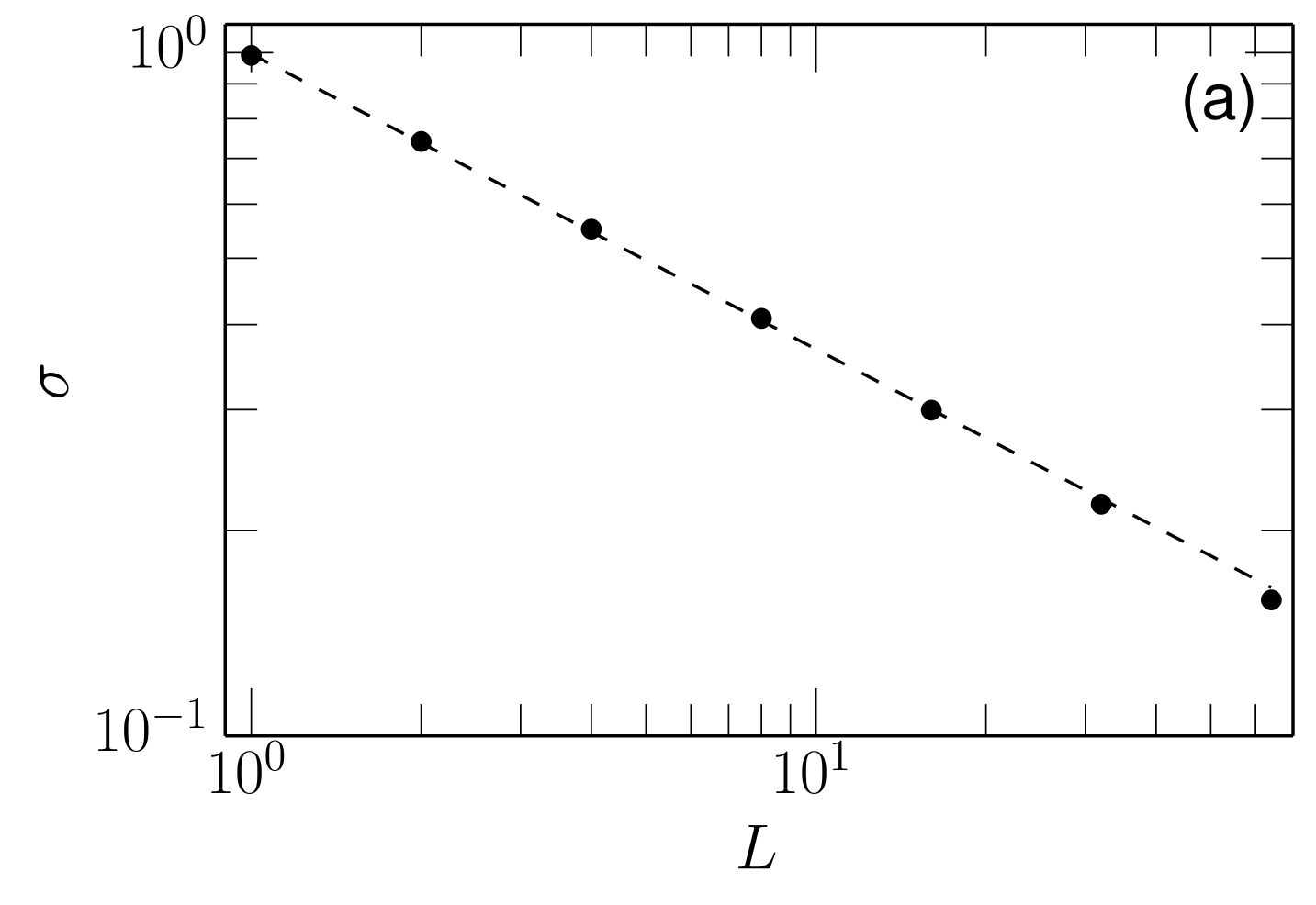}
\includegraphics[width=0.4\textwidth]{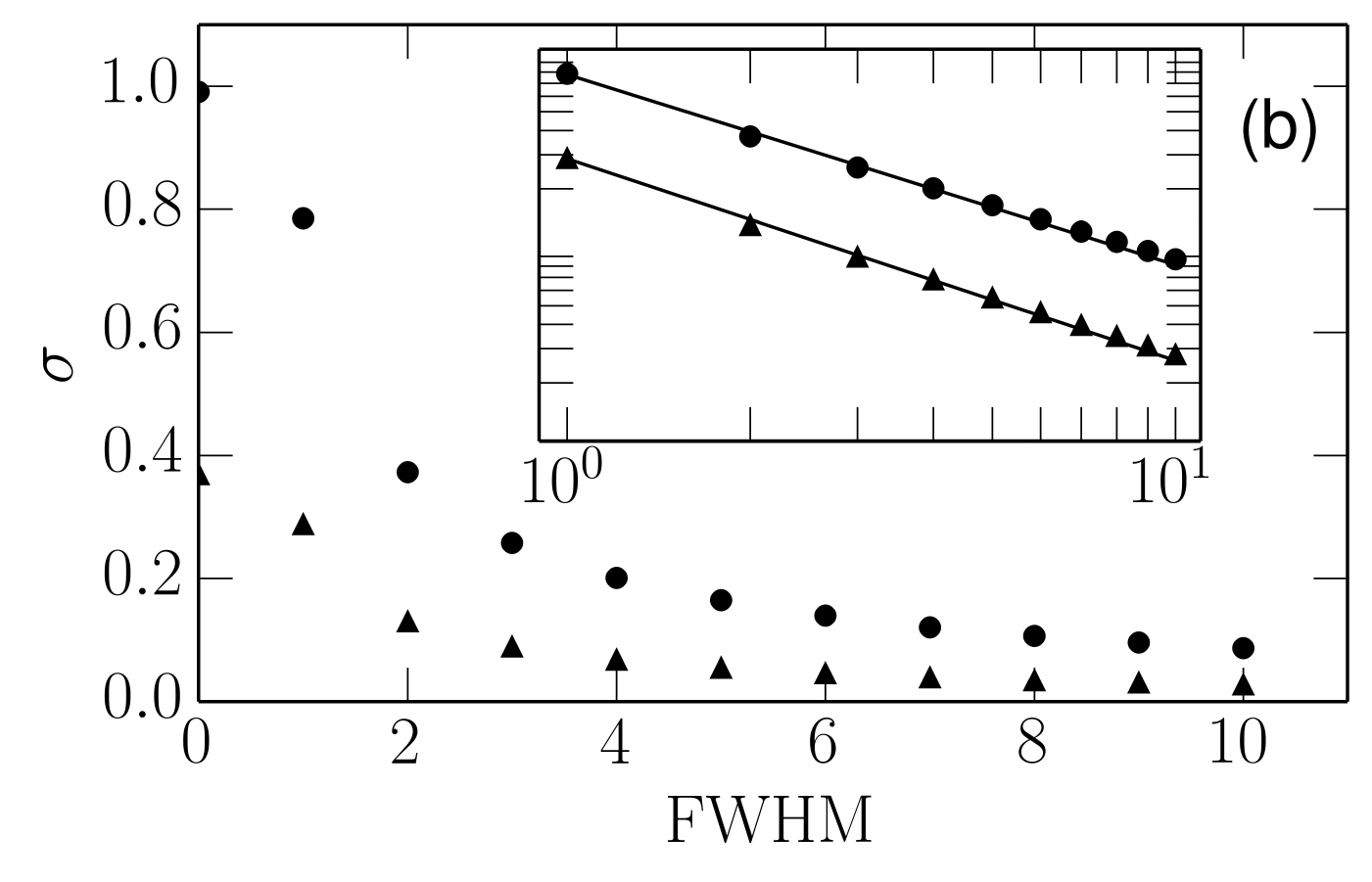}
\end{center}
\caption{(a) The effect of an increasing LOS on the dispersion 
($\sigma$ in Eq.~(\ref{eq:lognormal})) of the best fit lognormal PDF. The 
dashed line shows the best-fit power-law.  (b) The effect of smoothing with 
a 2D Gaussian, of unit mean and full-width half-maximum shown on the 
horizontal axis, on $\sigma$. Circles are for models with $L=1$ and triangles 
for $L=10$. The inset panel shows the best-fit power-laws to the data points 
where $\mathrm{FWHM}\ge 1$. In both panels the modelled column densities were 
calculated using the simple model described in Section~\ref{sec:models}.}
\label{fig:LOSBeam}
\end{figure}

%_____________FIGURE_5___________
\begin{figure}
\begin{center}
\includegraphics[width=0.4\textwidth]{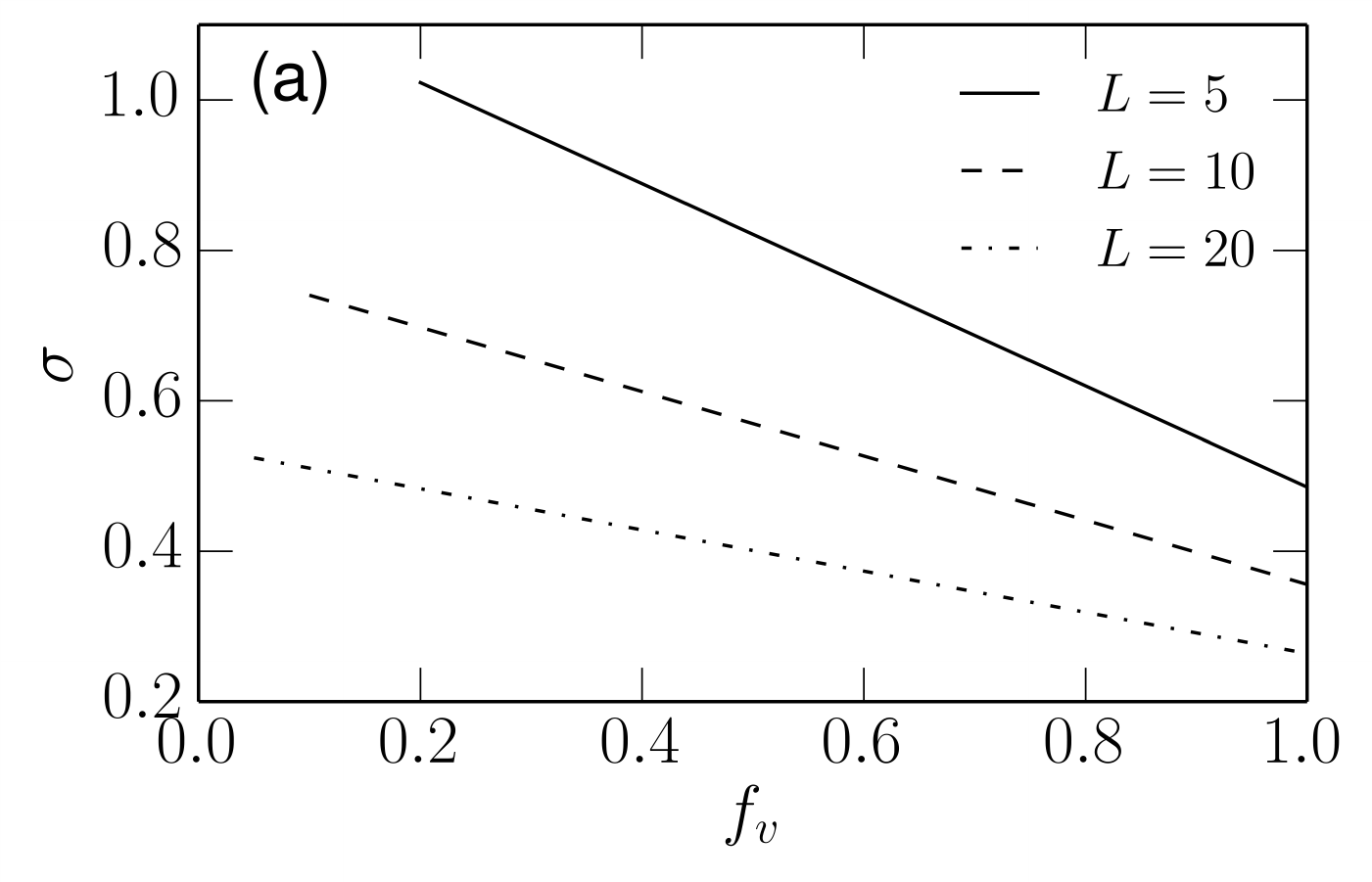}
\includegraphics[width=0.4\textwidth]{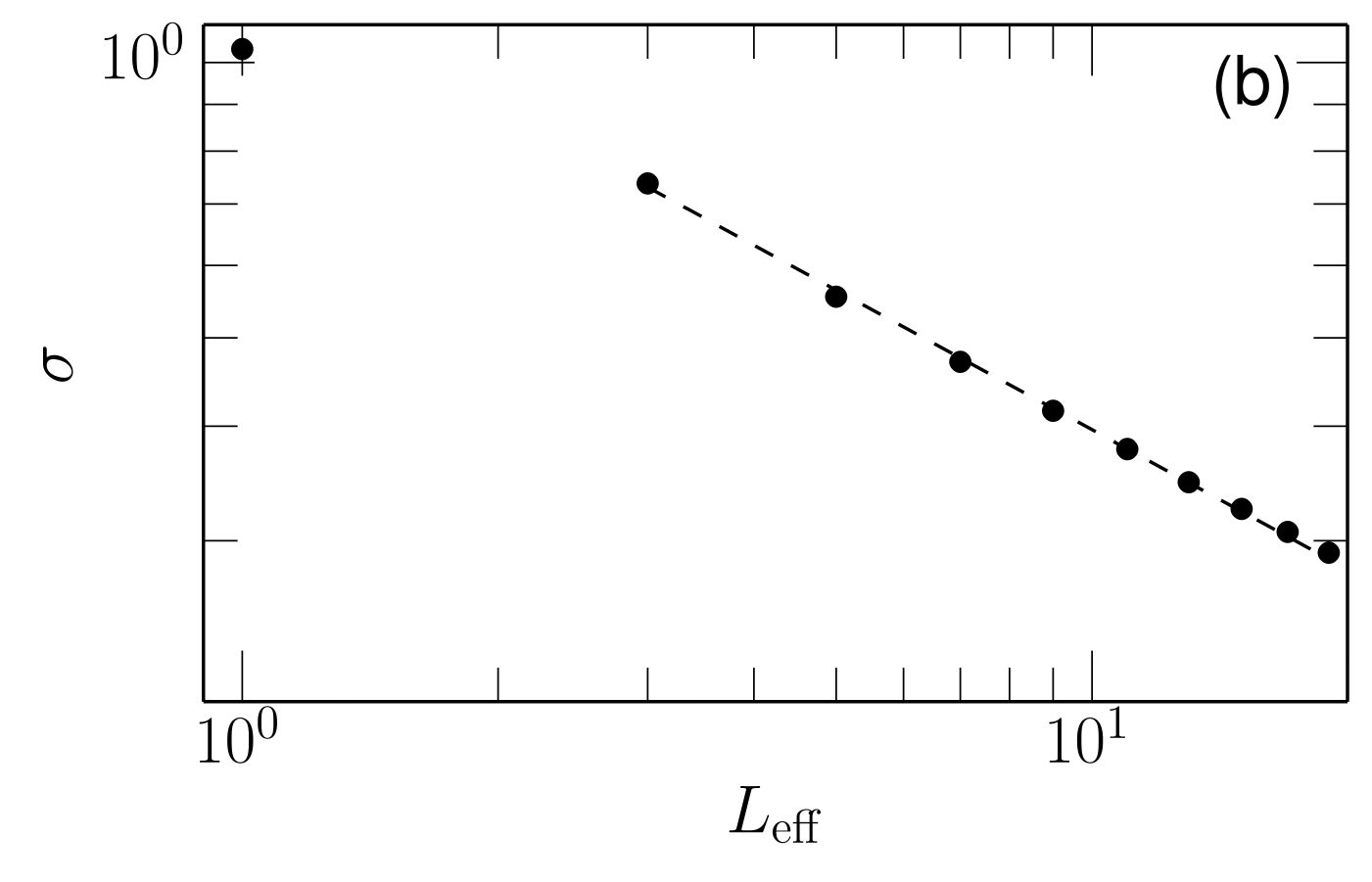}
\end{center}
\caption{(a) The effect of varying fractional volume, $f_{v}$, on the 
dispersion ($\sigma$ in Eq.~(\ref{eq:lognormal})) of the best fit lognormal 
PDF of the average volume density. Lines are shown
for different LOS. (b) $\sigma$ versus the effective LOS $\Leff$ as defined 
in the text. The best-fit power-law for $\Leff>1$ is shown. In both panels 
the modelled average volume densities, the column density divided by the LOS, 
were calculated using the simple model described in Section~\ref{sec:models}.}
\label{fig:fvLeff}
\end{figure}

The maps used to compute the column and average volume density histograms 
presented in Section~\ref{sec:PDFs} all have a resolution of a few hundred 
parsec. The presence of a lognormal density PDF does not in itself indicate
that the gas is turbulent, the lognormal PDFs are merely compatible with this assumption 
\citep{Tassis:2010, Hennebelle:2012}. One could consider two 
contributions to the column density PDFs: one from the 
density structure arising from compressible turbulence in the ISM and the 
other from the densities of specific, discrete objects that are not caused 
by turbulence, such as SNRs and \HII\ regions (both of which may, however, be drivers
of turbulence). How to separate these contributions to the PDF is not 
clear. Both types of structure have sizes less than the beam. The dominant 
energy input driving turbulence in the ISM is probably due to stellar cluster winds and 
superbubbles produced by clustered supernovae, with a typical scale in the 
range $50$--$500\,\pc$ \citep[][Section 3]{Elmegreen:2004}. Many other drivers of turbulence,
such as spiral shocks, the magneto-rotational instability and proto-stellar jets have also been proposed
\citep[see][for a discussion of these in the context of PDFs]{Federrath:2013b}. The \HH\ at the 
largest scale is collected in giant molecular clouds (GMCs), with a typical 
size of $50\,\pc$ \citep{Blitz:2007} to $100\,pc$ in M51 \citep{Colombo:2014},
and the \Ha\ has a diffuse component as well as higher density 
\HII\ regions of typical size $30\,\pc$ \citep{Habing:1979, Scoville:2001, 
Azimlu:2011}. The thickness of the diffuse gas layer is of the order of 
$500\,\pc$ and that of the \HH\ $150\,\pc$, so each telescope-beam cylinder 
will contain one or many turbulent cells or GMCs or \HII\ regions. 
Furthermore, the ISM consists of several phases, so the emission that we use 
to construct the average volume density PDFs is not uniformly distributed 
along the total LOS but only arises from a fraction of this distance. In 
order to understand how averaging over multiple cells or clouds, both along 
and across the beam, and how the filling factor of the ISM occupied by the 
phase we observe affect the density PDFs we constructed a simple model. 

\subsection{A simple model for the random gas density distribution in the ISM}

Independent random numbers drawn from a lognormal distribution with parameters 
$\mu=0$, $\sigma=1$ --- see Eqs.~(\ref{eq:lognormal}) and (\ref{eq:gaussian}) 
--- were distributed over a Cartesian $(x,y,z)$ grid of $(1000\times 1000\times
 L)$ mesh-points. The value of each mesh-point represents the gas density of a 
single gas cloud and the unit of distance in the model is thus the cloud size.
The parameter $L$ represents the line-of-sight and for $L>1$ the random numbers
 were summed along the $z$-axis to give a map of the column density of the 
model, with $1000\times 1000$ lines of sight. An average volume density along 
each line of sight was also computed by dividing the column density by $L$. 
Convolution of the column density with a 2D Gaussian of a given
full-width half-maximum (FWHM) was used to model the consequences of smoothing
by a beam of size $D=\mathrm{FWHM}$.

Making ($1-f_{v}$) of the mesh-points empty allowed us 
to control the fractional volume occupied by the clouds,
$f_{v}=V/(1000\times1000\times L)$, where $V$ is the number of meshpoints for 
which the gas density is non-zero. The empty mesh-points were chosen at random.
We shall refer to $f_{v}$ as the volume filling factor but stress 
that there are at least two other quantities that are also called filling
factors. Both measure the ``clumpiness'' of a given gas phase, either averaged 
over the entire volume or only the volume occupied by that phase, and 
calculated using the ratio $\langle n\rangle^{2}/\langle n^{2}\rangle$ where 
$n$ is the gas density and $\langle\dots\rangle$ represents the appropriate 
average: note that the relation between these filling factors and $f_{v}$ is 
not straightforward \citep[see][Sect.~5]{Gent:2013}. Under the assumption that
the gas is confined to clouds which all have the same density, which is commonly
made when deriving filling factors from observational data \citep[see 
e.g.][]{Berkhuijsen:2008}, the fractional volume occupied by the gas and the 
volume filling factor defined as $\langle n\rangle^{2}/\langle n^{2}\rangle$
are the same. We adopt this assumption in what follows; in other words we 
assume that the fractional volume and the volume filling factor are one and the
same quantity. 

Normalised histograms of the column or average-volume density gave the model 
PDFs, and the best-fitting lognormal function to the PDF, using a least-squares 
method, was used to compute $\mu$ and $\sigma$ for each model. Ten independent 
realisations were calculated for each set of model parameters and the resulting
$\mu$ and $\sigma$ averaged. For $L=1$, $f_{v}=1$, and perfect resolution (no 
smoothing), the model PDFs have the same parameters, $\mu=0$ and $\sigma=1$, as 
the lognormal distribution used to populate the model (within a numerical error
due to the finite size of the mesh and the constraints of the fitting 
algorithm). We used the model to explore how $\sigma$, as this is by far the 
most useful of the two parameters in interpreting observations, varies as $L$, 
$f_{v}$ and $D$ change.  

\subsection{PDF dispersion and the pathlength}
Figure~\ref{fig:LOSBeam}(a) shows how the dispersion of the lognormal PDF, 
$\sigma$, changes as the pathlength along the line of sight, $L$, is increased.
Here $f_v=1$ and no smoothing is applied.
For $L=1$ the model recovers $\sigma=1$, as required by the parameters of the  
original distribution of densities. As $L$ increases $\sigma$ decreases as a 
power law $\sigma\propto L^{\alpha}$. The fitted power law in Fig.~
\ref{fig:LOSBeam}(a) has $\alpha=-0.4$ for the range $1<L<64$, but as the lower 
end of the range in $L$ increases (i.e. for $L\gg1$) we find that $\alpha
\rightarrow -0.5$. This scaling is in agreement with \citet{Fischera:2004} who 
studied the relationship between $\sigma$ of the column density PDF and the 
ratio of $L$ to the size of a turbulent cell, $L/\turb$: for thick discs, i.e. 
$L/\turb \gg 1$, they derived $\sigma\propto (L/\turb)^{-0.5}$. In our simple 
models $\turb=1$ and so $L/\turb\ge 1$.

\subsection{PDFs of column and average volume densities}
The scaling of $\sigma$ with $L$ for the lognormal PDFs of the average volume 
density, defined as $\langle n\rangle=\Sigma/L$, follows the same power law 
behaviour as the column density PDFs: for $n$ we also find that $\sigma\propto 
L^{-0.5}$ for $L\gg 1$. This is not surprising, as dividing a lognormal by a 
constant changes $\mu$ but not $\sigma$ 
\citep[e.g.][Theorem 2.1]{Aitchison:1957}. 
In these models we set $f_v = 1$ and applied no smoothing.

\subsection{PDF dispersion and beam smoothing}
Figure~\ref{fig:LOSBeam}(b) shows how $\sigma$ changes under smoothing of the 
model column (and volume) densities, where we have fixed $f_v=1$. Smoothing 
reduces $\sigma$ and for beamwidths $D\ge 1$ the scaling follows a power law, 
shown in the inset to the Figure, with $\sigma\propto D^{-1}$ irrespective of 
$L$. This power law behaviour means that when $D\gtrsim 4$-$5$ clouds for 
shallow $L$ and $D\gtrsim 2$-$3$ clouds for longer $L$, the effect on $\sigma$ 
of further moderate smoothing is marginal: the resolution of observations of 
nearby galaxies is typically in the range $1<D<10$. In principle, if 
observations are available with a resolution of $D<\turb$ and repeated smoothing
results in $\sigma\propto D^{-1}$ beyond a certain beamwidth, one could recover 
$\turb$ from the column density PDFs. 

\subsection{PDF dispersion and the filling factor} 
Figure~\ref{fig:fvLeff}(a) shows that as the filling factor increases, i.e. as 
the gas occupies more of the total volume, the 
dispersion of the fitted lognormal PDFs for the average volume density 
decreases linearly. However, the 
effect on $\sigma$ depends on both $L$ and $f_{v}$. If $f_{v}$ is a constant 
for all lines of sight, then the power law scaling 
$\sigma\propto L^{{\alpha}}$ with $\alpha=-0.5$ still holds. However, if 
$f_{v}$ depends on $L$ this scaling can change. This motivates us to define the 
effective pathlength 
\begin{equation}
\Leff=f_{v}L
\label{eq:Leff}
\end{equation}
which is the fraction of the total line of sight that is occupied by 
clouds or turbulent cells. Figure~\ref{fig:fvLeff} shows that $\sigma$ scales 
as a power law $\sigma\propto\Leff^{\alpha}$ for $\Leff>1$. The best fit power 
law has an index $\alpha=-0.5$, as in the models where $f_{v}=1$ shown in 
Fig.~\ref{fig:LOSBeam}(a).

\subsection{Summary of model results}
For convenience we summarise the results obtained using this simple model for 
a lognormal distribution of gas density in the ISM. We shall use these results 
to interpret the observations.
\begin{enumerate}
\item The dispersions of the lognormal column density and average volume 
density PDFs both scale as a power law with the total line of sight, 
$\sigma\propto L^{\alpha}$. When the filling factor $f_{v}=1$ both have 
$\alpha=-0.5$.
\item Smoothing by the telescope beam, when the beam is bigger than the size 
of a cloud, results in a power law scaling $\sigma\propto D^{-1}$.
For $D\gtrsim 5\turb$ in a thin layer ($L/\turb\sim 1$) and $D\gtrsim 2\turb$ 
in a thicker layer ($L/\turb\sim 10$), further smoothing only produces a weak 
effect on $\sigma$.
\item $\sigma$ is linearly related to the filling factor of the gas but the 
magnitude of the effect also depends on $L$. Defining the effective pathlength 
$\Leff=f_{v}L$ results in the scaling $\sigma\propto\Leff^{-0.5}$.     
\end{enumerate}

%_____________________________SECTION  5___________________________________

\section{Discussion}
\label{sec:discussion}

Two differences between the PDFs shown in Figs.~
\ref{fig:M31}--\ref{fig:M31:ring} and Tables~
\ref{table:M31:surface}--\ref{table:M31:volume} are immediately
obvious: the M51 PDFs are significantly wider than the M31 PDFs and
for both galaxies the \HH\ PDF is wider than the \HI\ PDF. Which
factors could cause these differences?

Simulations of the ISM have shown that the dispersion of a PDF depends on 
variables like disk mass, temperature, magnetic field strength, length 
of the line of sight and density scale. Furthermore, we showed in
Section~\ref{sec:models} that also the filling factor plays a role. 
Below we consider each of these factors separately. 

The width of the PDF increases with the gas density in the disc
\citep{Tassis:2007, Wada:2007}. In the HD simulations of \citet{Wada:2007} an 
increase of a factor of 10 in disc mass leads to an increase in the dispersion 
of 30\%. As the mean face-on column density of the total gas in M51 is about 
5 times higher than that in M31 (calculated from the \HI\ and \HH\ PDFs with 
$\avg{N}=X_0\exp(\sigma_{\ln}^{2}/2)$ and added \citep{Ostriker:2001}), this 
could account for a difference of about 15 per cent in the dispersions of the 
gas PDFs for the two galaxies.  

In the MHD simulations of \citet{Avillez:2005}, the dispersion of the
PDF of cold gas is wider than that of warm or hot gas. This was 
confirmed by the HD simulations of \citet{Robertson:2008} and 
\citet{Gent:2013}. Since the molecular gas is generally cooler than the 
atomic gas, the temperature difference could cause the \HH\ PDF to be wider 
than the \HI\ PDF. Unfortunately, we cannot estimate the expected difference 
in dispersion because the temperature ranges in the figures of these 
simulation papers are too large.

\citet{Molina:2012} simulated the development 
of molecular clouds with time using a MHD code. They derived relations 
between the dispersion of the PDF, the Mach number $M$ and plasma $\beta$, 
the ratio between thermal and magnetic pressure. They found that increasing 
$M$, i.e.\, more cooler and denser gas, widens the PDFs of the clouds 
\citep[see also][]{Federrath:2008, Price:2011, Federrath:2013a}, consistent with our 
observations, whereas decreasing 
$\beta$  makes the PDFs somewhat narrower. We can estimate the ratio of 
mean $\beta$ in M51 to that in M31 by comparing the thermal and
magnetic pressures in the two galaxies. As the total thermal pressure is 
dominated by the ionized gas, we use $X_0$ of the \Ha\ PDFs (Tables 1 and 2)
and the lines of sight $L$ given in Tables 4 and 5 to calculate the rms
electron density $\sqrt{\avg{n_e^2}}$ for each galaxy. This is 2.9 times higher in
M51 than in M31 suggesting that in M51 the thermal pressure is about three
times higher than in M31. As the magnetic field in M51 \citep{Fletcher:2011}
is about three times  stronger than in M31 \citep{Fletcher:2004}, the 
magnetic pressure in M51 is about nine times higher than in M31; hence, $\beta$
in M51 is about one third of that in M31. \citet{Molina:2012} set
$\beta \propto M^{-2}$ in their simulations of molecular clouds. If this
relationship also holds in less dense gas phases, a three times smaller 
$\beta$ would imply a $\sqrt{3}=1.7$ times higher mean $M$ in M51 than in M31.  
Since $M$ affects the PDF dispersion more strongly than $\beta$ 
\citep[see][]{Molina:2012}, the PDFs for M51 will be wider than those for M31,
in agreement with our observations. Separate measurements of $M$ and $\beta$
will be necessary to see how large this effect is compared to other influences
on the width of the PDF.

The scale of the density fluctuations influences the dispersion of the PDF 
because an increase in the number of clouds seen by the radio beam  
leads to a decrease in the dispersion as extremes are averaged out \citep
{Ostriker:2001,Vazquez-Semadeni:2001}. In Section~\ref{sec:models} we 
demonstrated that this effect may occur across the beam area as well as along 
the LOS. In Section~\ref{subsec:beam} we show that the angular resolution 
hardly influences the observed dispersion, and in Section~\ref{subsec:LOS} 
that the effective length of the LOS is the dominant factor in shaping 
the observed PDFs.

%________________________SECTION   5.1______________________________

\subsection{Does the observed dispersion depend on beamwidth?}
\label{subsec:beam}

In Section~\ref{sec:models} we showed that the dispersion $\sigma$ of the 
column density PDFs varies strongly under smoothing when the smoothing 
beamwidth is smaller than about $5$ clouds, but that once the beamwidth is 
larger than this the effect of further smoothing becomes negligible.  
Thus the behaviour of the best fit $\sigma$ under repeated smoothing can, 
in principle, be used to set a rough upper limit on the correlation size of 
the clouds. 

%_____________FIGURE_6___________
\begin{figure}
\begin{center}
\includegraphics[width=0.45\textwidth]{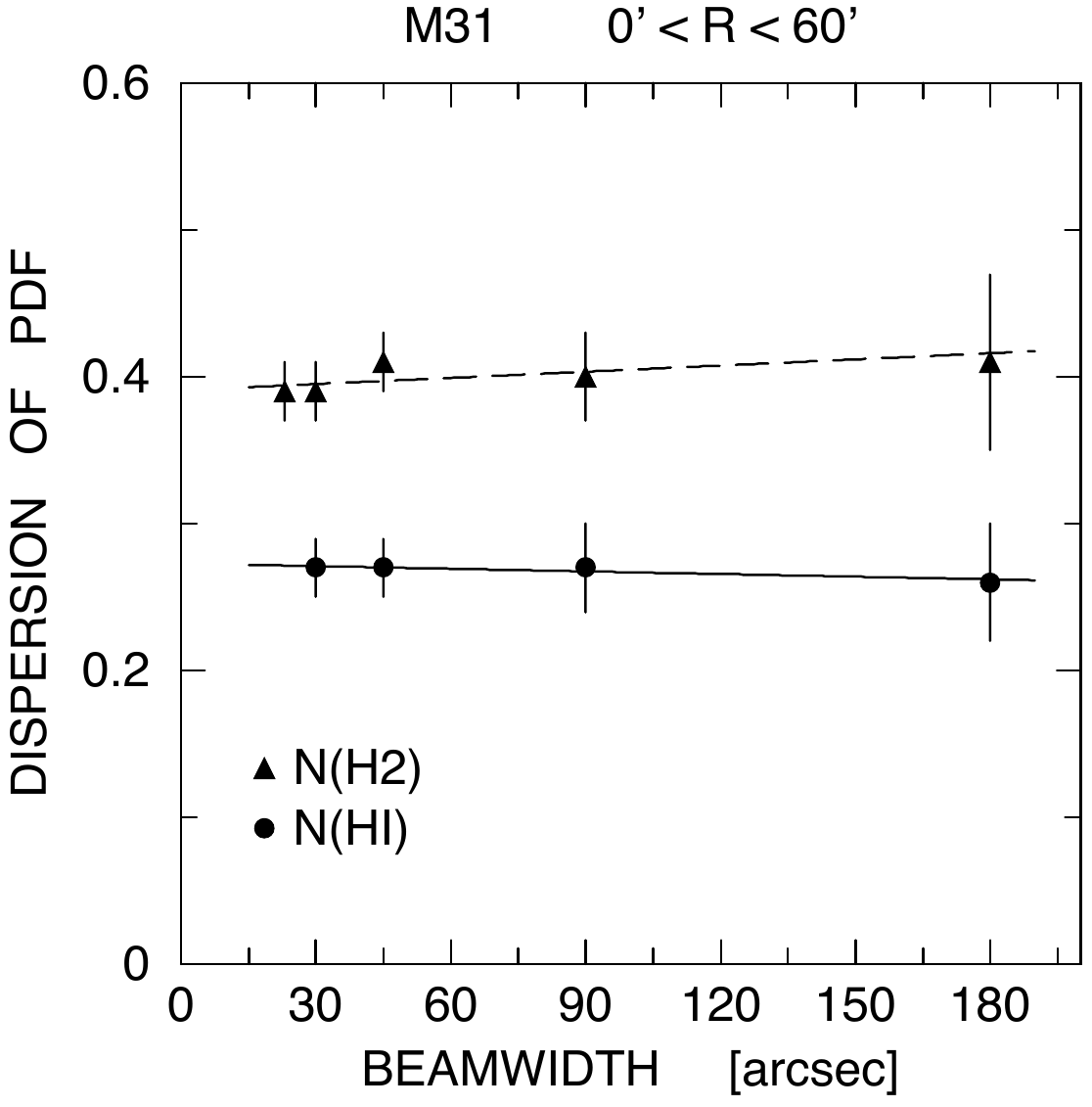}
\end{center}
\caption{Dependence of the dispersion $\sigma$ of the PDFs of the column densities
\NHI\ (dots) and \NHH\ (triangles) on the half-power beamwidth (in arcseconds) 
for the radial interval $0\arcmin<R<60\arcmin$ in M31. A beamwidth of 
$45\arcsec$ corresponds to $170\,\pc \times 785\,\pc$ in the plane of the 
galaxy. The lines are weighted least-squares fits to the data points.}
\label{fig:M31:disp}
\end{figure}

We investigated this point for the column densities of \HI\ and \HH\ of M31. 
To this end we smoothed the \HI\ and \HH\ maps with Gaussian functions
to angular resolutions of 24\,arcsec $\times$ 36\,arcsec (indicated by 
30\,arcsec), 90\,arcsec and 180\,arcsec, and calculated the PDFs. Including 
the PDFs at 45\,arcsec and the original resolution of 23\,arcsec of the 
\HH\ map, the range spans a factor of 7.8 in scale. We found that for the full
region $R<60\arcmin$ as well as for the subregions the dispersions of the PDFs
are the same within the errors for both \HI\ and \HH\ 
(see Fig.~\ref{fig:M31:disp}). Hence, the dispersion of the gas PDFs for M31 
is independent of smoothing scale between about $90\,\pc \times 400\,\pc$ and 
$700\,\pc \times 3150\,\pc$ in the plane of M31. 

Since the LOS through the disc of M31 is large (on average $~3000\,\pc$ for the 
\HI\ disc), Fig.~\ref{fig:LOSBeam}(b) suggests that smoothing will not affect 
$\sigma$ when the FWHM is wider than about 2 clouds. We can use this to make 
an estimate of the upper limit to the size of the clouds in M31. The smallest 
of the smoothing scales we consider, $90\,\pc \times 400\,\pc$, must contain at 
least $2^2=4$ clouds, giving an estimate for the upper limit of 
the size of the clouds of $\turb<\sqrt{(400\times 90)/4}=95\,\pc$.

Note that the linear resolution of the M51 observations is about $300\,\pc$, 
similar to that of the best resolved \HI\ observations of M31. Therefore, the 
difference in dispersion of the PDFs of the gas surface densities for M31 and 
M51 cannot be caused by a difference in angular resolution, provided the 
scales of the clouds are the same.

%_____________________________________SECTION   5.2_____________________________

\subsection{Line-of-sight effects}
\label{subsec:LOS}

In the case of column density PDFs, Figure~\ref{fig:LOSBeam}(a) shows that at 
a fixed resolution the fitted $\sigma$ will be smaller by about a factor 
of $2$ for a $4$-times longer LOS. In particular, we found that $\sigma$ 
decreases with $L^{-0.5}$. We observe this effect in the PDFs of M31 and 
M51 in several ways.

First, the PDFs of M31 (Fig.~\ref{fig:M31}) have a smaller dispersion
than those of M51 (Fig.~\ref{fig:M51}). The LOS through the strongly
inclined galaxy M31 is typically about 4 times longer than the LOS
through M51 that is nearly seen face-on, provided that the scale
heights are similar. Assuming that the cloud scales in the two
galaxies are comparable, we then expect more clouds along the
LOS through M31 than along the LOS through M51, and hence smaller
dispersions of the PDFs for M31 than for M51.

Second, for both M31 and M51 the PDFs of \HH\ are significantly wider
than those of \HI. Since the molecular gas is more concentrated in the disc
mid-plane than \HI\ (due to the larger gravitational forces) and  
molecular clouds are generally denser and smaller than \HI\ clouds, the LOS 
through molecular gas are shorter than through \HI\ gas leading to a larger 
dispersion of the \HH\ PDF than of the \HI\ PDF. Also the Mach number 
of the \HH\ gas is higher than of the \HI\ gas (because the temperature is much 
lower whereas the turbulent velocity is similar), causing a higher compression 
of the gas in the molecular clouds by  turbulent shocks and larger dispersions 
of the cloud PDFs (\citet{Molina:2012} and references therein). However, this 
effect may be reduced if the LOS intersects many clouds, or if the beamwidth 
is larger than the typical cloud size: in this case the peak of the PDF may 
move to a higher density but the dispersion can be unchanged. Which of these 
two effects is responsible for the observed difference in the PDF dispersions 
requires further study. 

Third, the \HI\ PDF for the inner region of M31 is slightly wider
($\sigma=0.23\pm 0.01$) than that for the bright 'ring'
($\sigma=0.19\pm 0.01$) where the LOS is longer and goes through more
clouds than in the inner region. This interpretation is
supported by the higher value of $X_0$ in the 'ring'. The same trend
is visible in the \HH\ PDFs (see Table~\ref{table:M31:volume}).

Furthermore, the dispersions of gas PDFs derived in numerical
simulations of the ISM \citep[e.g.][]{Vazquez-Semadeni:2001,
Avillez:2005, Wada:2007} are usually larger than observed PDFs.
\citet{Berkhuijsen:2008} attributed the difference to the higher
densities used in the simulations compared to observed densities.
However, the much smaller lines of sight available in simulations than
in observations may be the main cause of the larger dispersions of
PDFs derived from simulations \citep[see also][]{Fischera:2004}.

It is possible to quantify the relationship between the dispersion of
column density PDFs and LOS. 
Because the scale height of the gas in M31 
increases outwards, we can test this relationship using the PDFs of the 
average volume densities of the sub-regions at different radii given in 
Table~\ref{table:M31:volume}.

%_____________FIGURE_7___________
\begin{figure}
\begin{center}
\includegraphics[width=0.45\textwidth]{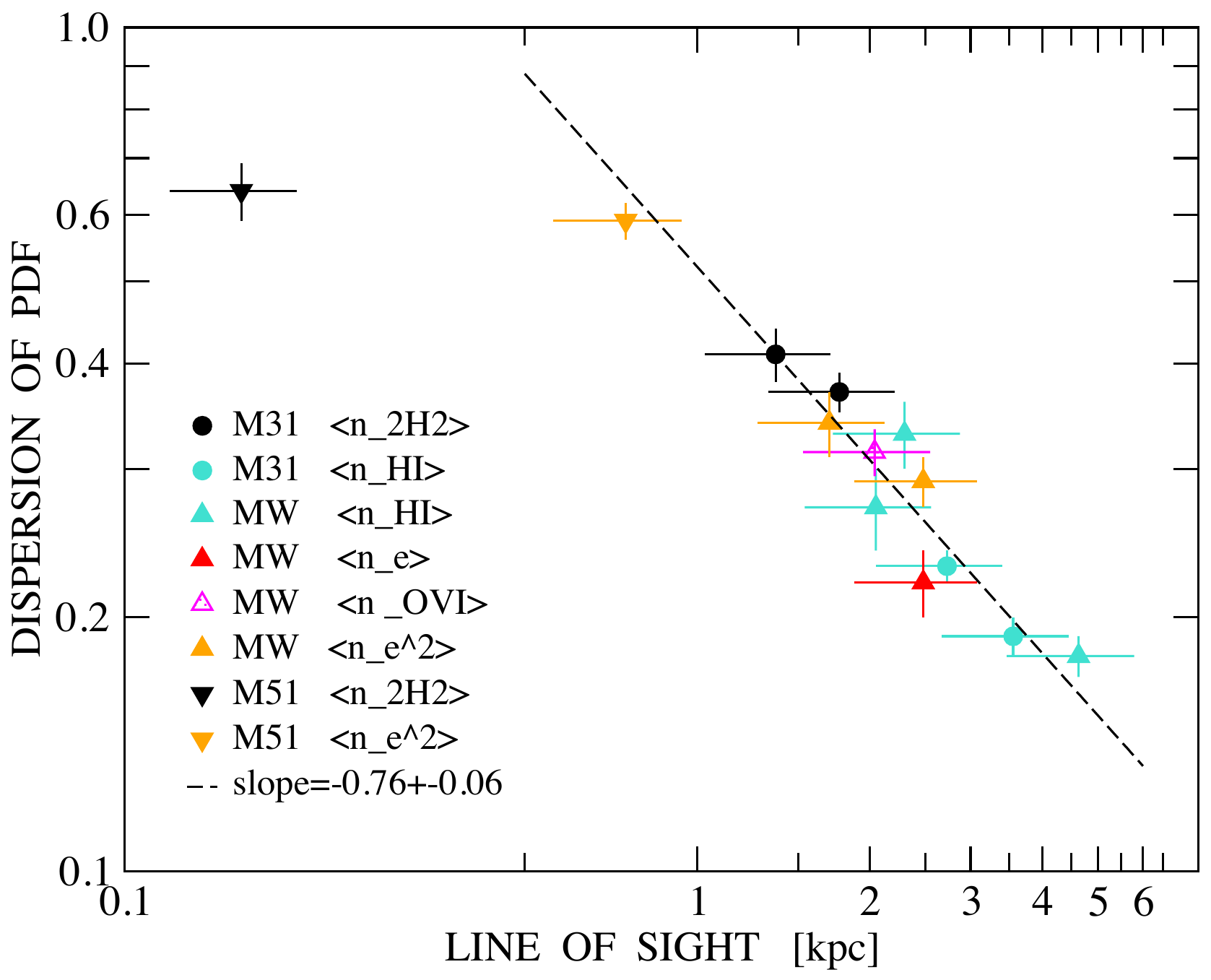}
\end{center}
\caption{Dependence of the dispersion $\sigma$ of the PDFs of mean gas volume 
densities in M31, the Milky Way and M51 on line of sight $L$ (in pc).
We assumed an uncertainty of $25\%$ in $L$. 
The dashed line is the weighted least-squares fit given by Eq.~\ref{eq:los}.
(The figure is shown in colour in the electronic version of the paper.)}
\label{fig:los}
\end{figure}

In Fig.~\ref{fig:los} we show the dispersion of the PDFs of average 
volume density of \HI\ and 2\HH\ as function of the mean LOS $L$ of the 
radial ranges $0\arcmin<R<35\arcmin$ and $35\arcmin<R<60\arcmin$, where
$L=2h(R)/\cos{i}$ and $h(R)$ is the scale height as derived in
Section~\ref{subsec:volume}. 

We have added 3 points for the mean volume density of diffuse \HI\ (cool 
and warm \HI) towards stars in the solar neighbourhood (SN) and for the mean 
volume density of free electrons in the diffuse ionized gas (DIG) at Galactic 
latitudes $|b|>5\degree$ within a few kpc from the sun as well as two
points of mean electron density squared obtained from emission measures towards
pulsars, all taken from 
\citet{Berkhuijsen:2008, Berkhuijsen:2012}. We also show the dispersion of the
\OVI\ PDF calculated along the lines of sight to 190 stars using fig. 9 (left 
panels) in \citet{Bowen:2008}. 

For M51 scale heights of molecular and ionized gas are available.
\citet{Pety:2013} combined CO surveys observed with Plateau de Bure and  the
30-m IRAM telescope. An extensive analysis showed that the molecular gas not 
only has a thin disc with a Gaussian scale height of $h=40\,\pc$, but also a 
thick disc with a scale height of about 250\,\pc.
\footnote{We use the upper value of the range 190-250\,\pc\, given by 
\citet{Pety:2013} because it applies to the larger radii covered by our 
data.} Taking the about 8 times higher volume density of the thin disc in the 
mid-plane into account, their sum can be desribed by a single exponential with 
a scale height of about 75\,\pc. \citet{Seon:2009} modelled the \Ha\ emission 
from M51 observed by \citet{Thilker:2000} and derived a thin disc of \HII-
regions with exponential scale height of 300\,\pc\, and a thick disc of DIG 
with 1\,\kpc\, scale height. In the mid-plane the emission from the thin disc 
is about 7 times stronger than that from the DIG. The sum of the two 
distributions can be approximated by a single exponential disc with a scale 
height of 350\,\pc. We have added the dispersions of \NHH\ and \Ha\ given in 
Table~\ref{table:M51} using the single scale heights obtained above for the 
LOS $L$. Details on the data points in Fig.~\ref{fig:los} are given in 
Table~\ref{table:los}. 

Apart from the \NHH-point of M51 (that we discuss below), the dispersions
in Fig.~\ref{fig:los} systematically decrease with increasing $L$. A weighted
fit through the 12 available points yields the power law 
\begin{equation}
\log[\sigma(L)]=(2.0\pm1.7)-(0.76\pm0.06)\log[L]
\label{eq:los}
\end{equation}
with $\chi^2=0.3$ and $L$ in pc.

We conclude that the length of the line of sight through the gas in
M31, the MW and M51 is the main influence on the width of the PDFs of 
average volume density. This implies that the number of clouds along the LOS 
determines the dispersion of the PDF.

Since the data points for the Milky Way and \Ha\ for M51 in Fig.~\ref{fig:los} 
are consistent with the M31 points, the variation in cloud sizes must be 
similar in these galaxies. This conclusion is supported by the work 
of \citet{Azimlu:2011}, who found that the size distributions of \HII\ regions 
in different galaxies are very similar, but the total number of \HII\ regions 
differs depending on the star formation rate.

We note that most of the data points in Fig.~\ref{fig:los} forming the
$\sigma$ - $L$ relation refer to diffuse gas. Much of the \HI\ in M31 is known
to be diffuse \citep{Braun:1992} and the small amount of molecular 
gas in M31 may 'look' diffuse in our large radio beam because it does not 
contain many small, dense \HH\ clouds.
 
%-----------------------TABLE 4  CORRECTED--AND  EXTENDED ----\ref{table:los}----------------
\begin{table*}
\caption{Effective lines of sight $\Leff$ and filling factors}
\begin{tabular}{lllrrrrrr}
\hline
Gal. & Gas & Zone & N & L & $\Leff$ [pc] & $f_v$  & Dispersion & Ref. \\
\hline
M31 & $2\HH$ & $R<35\arcmin$ & 248 & 1370 &  $60\pm15$ & $0.04\pm0.02$ & $0.41\pm0.03$ & 1 \\
    & &        $R>35\arcmin$ & 672 & 1770 &  $70\pm15$ & $0.04\pm0.01$ & $0.37\pm0.02$ & 1 \\
    & $\HI$  & $R<35\arcmin$ & 459 & 2730 & $180\pm40$ & $0.07\pm0.02$ & $0.23\pm0.01$ & 1 \\
    &     &    $R>35\arcmin$ & 998 & 3560 & $270\pm60$ & $0.08\pm0.02$ & $0.19\pm0.01$ & 1 \\
 & & & & & & & & \\
MW & $\HI_c$ & $|b|>5\degree$ &  64 & 2050 & $130\pm40$ & $0.06\pm0.03$ & $0.27\pm0.03$ & 3 \\  
   & $\HI_w$ & $|b|>5\degree$ &  98 & 2300 & $ 90\pm25$ & $0.04\pm0.01$ & $0.33\pm0.03$ & 2 \\    
   &         & $|b|<5\degree$ &  42 & 4630 & $300\pm65$  & $0.06\pm0.02$ & $0.18\pm0.01$ & 2 \\    
   & $N_e$   & $|b|>5\degree$ &  34 & 2480 & $200\pm50^{*}$ & $0.08\pm0.02$ & $0.22\pm0.02^{*}$ & 2 \\
   & $\OVI$  & $|z|>200\pc$   & 190 & 2040 & $100\pm25$ & $0.05\pm0.02$ & $0.31\pm0.02$ & 4 \\
   & $N_e^{2}$ & $|b|>5\degree$ &  34 & 2480 & $115\pm25$ & $0.05\pm0.02$ & $0.29\pm0.02$ & 2 \\
   &           &                & 157 & 1700 & $ 85\pm20$ & $0.05\pm0.02$ & $0.34\pm0.03$ & 2 \\
 & & & & & & & & \\
M51 & $\HH$   & $R<3\arcmin$   & 282 &  160 & $ 24\pm 6$ & $0.15\pm0.05$ & $0.64\pm0.05$ & 1 \\
    & $\Ha$    &                & 728 &  750 & $ 28\pm 6$ & $0.04\pm0.01$ & $0.59\pm0.03$ & 5 \\ 
\hline\noalign\\
\end{tabular}
\medskip
         
N=number of LOS; assumed errors in L: 25\%, 
$^{*}$ $\sigma=0.22\pm0.02$ and $\Leff=200\pm50\,\pc$ calibrate $\Leff$ (see 
text). References: 1. This work; 2. Berkhuijsen \& Fletcher (2008); 3. 
Berkhuijsen \& Fletcher (2012); 4. Obtained from data in Bowen et al. (2008); 
5. Seon (2009).
\label{table:los}
\end{table*}

%________________________________SECTION   5.3________________________________

\subsection{Importance of volume filling factor}
\label{subsec:fv}

The power-law dependence of $\sigma$ on $L$ presented in Fig.~\ref{fig:los} 
has an exponent of $-0.76\pm0.06$, which is $-0.26$ smaller than 
expected from the models in Sect.~\ref{sec:models}. However, the density 
structure in the ISM contains holes and voids in and between gas clouds  
\citep[e.g.][]{Elmegreen:1997, Elmegreen:1998, Avillez:2005} and so we should 
also take into account the volume filling factor, $f_v$, of the gas using 
Eq.~(\ref{eq:Leff}).

Estimates of $f_v$ from observations \citep{Dickey:1993, Berkhuijsen:1999} 
and simulations \citep{Avillez:2005, Gent:2013} have shown that \HI\ gas 
has a larger filling factor than the denser molecular gas. So in 
Fig.~\ref{fig:los} $f_v$ will increase with increasing $L$ and $\Leff$ will 
increase faster than $L$.

For the $\avg{n_e}$-point for the MW at $L=2480\,\pc$ in Fig.~\ref{fig:los}, 
\citet{Berkhuijsen:2008b} derived a volume filling factor of 
$f_v=0.08\pm 0.02$. Hence the effective LOS  producing the dispersion of 
$0.22\pm0.02$ is $\Leff=(0.08\pm 0.02)*2480 \approx 200\pm 50\,\pc$. 
Assuming that the relation $\sigma\propto \Leff^{-0.5}$ found in 
Sect.~\ref{sec:models} holds in the ISM, we can estimate \Leff\, for the other 
points in Fig.~\ref{fig:los} using the observed $\sigma$ in 
$\sigma/(0.22\pm 0.02)=(\Leff/(200\pm 50))^{-0.5}$. The volume filling factor 
then is $f_v=\Leff/L$.

We list the observed and derived properties of the 13 points in 
Fig.~\ref{fig:los} in Table~\ref{table:los}. The errors given are random errors
only. The error of 25 per cent in \Leff\, of the calibration point is 
systematic and affects all \Leff\, and $f_v$ in the same way. As expected, 
the values of \Leff\, increase with increasing $L$ varying between about 
$25\,\pc$ in M51 and more than $200\,\pc$ in the nearly edge-on galaxy M31. 
The filling factors, though, vary much less: in M31 between $0.04\pm0.01$ for 
\HH\  and $0.08\pm0.02$ for the \HI\ point at $L=3560\,\pc$. Apart from the 
deviating \HH\ point for M51, also the other filling factors are in this range. 
These values seem amazingly small, but they compare well with other information. 
For the cold (\HH\ and HI) gas at temperatures $T<200\K$, \citet
{Breitschwerdt:2005} found from their MHD simulations $f_v=0.06$. \citet
{Gent:2013} derived from their simulations $f_v=0.03$ for cold gas at 
temperatures $T<500\K$. In their OVI-line study, \citet{Bowen:2008} estimate 
that the size of the gas regions causing these lines is about 100-200\.\pc. 
Since the mean distance is $L=2040\,\pc$, $f_v=0.049-0.098$ for these regions, 
consistent with our value of $f_v=0.05\pm0.02$. We note that at $L=2480\,\pc$ 
$\sigma$ of $<n_e^{2}>$ is larger than $\sigma$ of $<n_e>$, as one would 
expect. The larger $\sigma$ of $<n_e^{2}>$ leads to values of \Leff\, and $f_v$ 
that are nearly half of those of $<n_e>$, in agreement with the smaller scale 
height and larger clumpiness of HII regions compared to the DIG.  

In Fig.~\ref{fig:los} the \HH\ point for M51 is located at a much smaller LOS 
than is expected from the relation formed by the other 12 data points. This 
may be due to the large filling factor of $f_v=0.15\pm0.05$, which is three 
to four times larger than $f_v$ of the \HH\ points for M31 (see Table~\ref
{table:los}). In the inner parts of M51 ($R < 3\arcmin$) the dominant gas 
component is \HH\  \citep[e.g.][]{Pety:2013}), while in M31 \HI\ is the dominant 
component \citep[e.g.][]{Nieten:2006}. At the end of Sect.~\ref{subsec:surface} 
we showed that the mean face-on column density of \HH\ in M51 is 11 times 
higher than that of M31, which explains the higher value of $f_v$ in M51. 
When the filling factor is high, already a small effective LOS contains so 
many clouds that the PDF becomes lognormal. As \Leff\, in M51 is only about 
$25\,\pc$, the maximum size of the clouds along the LOS must be much less than 
$25\,\pc$. 

Since the dispersion of the column density PDF is the same as that of the PDF 
of the average volume density (see Sect.~\ref{sec:models}), we can estimate 
$L$ of \Ha\ in M31 and of \HI\ in M51 from Eq.~\ref{eq:los} using the observed 
dispersions. This yields $L=1750\pm1500\,\pc$ for \Ha\ in M31, similar to that 
of \HH\ at $R>35\arcmin$. Using the same scaling for \Leff\, as above, we find 
$\Leff=85\pm20\,\pc$ and $f_v=0.05\pm0.02$ (see Table~\ref{table:losM31M51}). 
The value of $L$ implies a scale height of $190\pm160\,\pc$, the same as for 
the \HH\ in the emission ring. This scale height of \Ha\ is consistent with 
the observation of \citet{Walterbos:1994} that the z-extent of the DIG in 
M31 is less than about 500\,\pc. If Eq.~\ref{eq:los} and the scaling for 
$\Leff$ also hold for the \HI\ gas in M51, then the observed dispersion in 
Table~\ref{table:M51} indicates a LOS of about $1\,\kpc$ (see Table~
\ref{table:losM31M51}). The corresponding \Leff\, and filling factor of 
$\Leff=37\pm10\,\pc$ and $f_v=0.04\pm0.01$, respectively, are smaller than in 
M31, which may be due to the small \HI\ content in M51.  

%-------TABLE 5-------------\ref{table:M51:los}  RE-ARRANGED

\begin{table*}
\caption{Estimates of LOS from dispersion of column density PDF.}
\begin{tabular}{llcccc}
\hline
Gal. & Comp. & L  & $\Leff$ & $f_v$ &   Dispersion \\
 & & [pc] & [pc] &  & \\
\hline
M31    &  $\Ha$  &   $1750\pm 1500$ &  $85\pm 20$ &   $0.05\pm 0.04$&    $0.34\pm 0.02$ \\
 & $R<60\arcmin$ & & & &  \\
 & & & & & \\
M51     &  $\HI$ &   $1030\pm 900$ &  $37\pm 10$ &   $0.04\pm 0.03$ &    $0.51\pm 0.05$ \\
\hline\noalign\\    
\end{tabular}
\medskip

\label{table:losM31M51}
\end{table*}

Interestingly, \citet{Seon:2009} presented separate \Ha\ PDFs for the 
DIG and the \HII\ regions in M51 based on the data of \citet{Thilker:2000}. 
Although the scale heights differ by a factor of three (300\,\pc\, and 
1000\,\pc\, for \HII\ regions and the DIG, repsectively), both PDFs have a 
dispersion of $0.36\pm0.03$, but the position of their maxima differs by about 
a factor of 7. The LOS through the DIG would be $2130\,\pc$ and the dispersion 
of $0.36\pm0.03$ is not far from the value of $0.29\pm0.09$ expected from Eq.~
\ref{eq:los}, but the point for the $\HII$ regions PDF with an LOS of 
$640\,\pc\,$ disagrees with the value of $\sigma=0.73\pm0.14$ expected from Eq.~
(\ref{eq:los}). The same dispersion for the two layers indicates that $\Leff$ 
is the same and the difference in LOS then means that the volume filling 
factors and/or the maximum cloud scales differ. Using the same scaling as 
before, we find that $\sigma=0.36\pm0.03$ corresponds to $\Leff=75\pm20\,\pc\,$ 
giving $f_v=0.04\pm0.01$ for the DIG and $f_v=0.12\pm0.04$ for 
the disc of \HII\  regions. The filling factor of the \HII\ regions is similar 
to that of the \HH\ clouds in M51 suggesting that their density structure is 
similar.

It is worth noting that as the PDFs for the disc and the DIG in M51 are 
strongly shifted in $\log(X_0)$, their combination yields a much wider PDF 
than that of the separate components. Thus our \Ha\ PDF for M51 is not 
representative for small regions or specific components. This poses a 
general problem for the interpretation of PDFs for large regions in 
galaxies containing diffuse, low-density regions as well as dense spiral arms.

%_____________________________________SECTION  5.4__________________________

\subsection{Estimating Mach numbers using PDFs: a note of caution}    
\label{subsec:mach}
A connection between the dispersion of the density PDF $\sigma$ and the Mach 
number $\mathcal{M}$ of interstellar turbulence has been identified 
theoretically \citep[e.g.][]{Padoan:1997, Passot:1998, Ostriker:2001, Federrath:2008, Federrath:2010,
Price:2011, Molina:2012} and 
used to make an estimate of $\mathcal{M}$ from observations 
\citep[e.g.][]{Padoan:1997,Hill:2008, Berkhuijsen:2008}. However, since the 
magnitude of $\sigma$ depends on the length of the LOS $L$, the telescope 
resolution $D$ and the volume filling factor of the gas $f_{v}$, caution 
should be exercised in applying the theoretical relation to real data. To
make a reliable estimate of $\mathcal{M}$ one needs to know something about 
the distribution of the observed gas and so correct the observed $\sigma$ for 
the effects of $L$, $D$ and $f_{v}$. 

%___________________________SECTION  5.5 ___________________________________

\subsection{M31: PDFs of dust emission and extinction}
\label{subsec:M31}

In view of the generally lognormal shapes of the gas PDFs, one may ask
whether other constituents of the ISM also have lognormal PDFs. The
dust emission is a good candidate since dust optical depth and gas
column density are well correlated in the solar neighbourhood
\citep{Bohlin:1978} and in M31 \citep{Tabatabaei:2010}, and lognormal
PDFs of molecular gas column density and of extinction through dust clouds 
in the MW have been reported 
\citep[e.g.][]{Padoan:1997, Goodman:2009, Kainulainen:2009, Schneider:2013}.

%_____________TABLE__6__________

\begin{table*}
\caption{Lognormal fits to the PDFs of dust emission from M31, HPBW $45\arcsec$. The fitted function is 
$Y=(\sqrt{2\pi}\sigma)^{-1}\exp[-(\log_{10} X - \mu)^2/2\sigma^2]$.}
%\begin{center}
\begin{tabular}{llrrrrrr}
\hline
 & & & & \multicolumn{2}{c}{Position of maximum} & Dispersion & \\
$X$ & Units & $R(\arcmin)$ & $N$ & $\mu$ & $X_0$ & $\sigma$ & $\chi^2$  \\
\hline
$24\um$ & MJy/sr & 0--60 & 
	1508 & $-0.47\pm0.01$ & $0.34\pm0.01$ & $0.34\pm0.01$ & 1.2 \\
 & & 0--35 & 525 & $-0.48\pm0.02$ & $0.33\pm0.01$ & $0.33\pm0.01$ & 1.4 \\
 & & 35--60 & 983 & $-0.46\pm0.01$ & $0.35\pm0.01$ & $0.35\pm0.01$ & 0.8 \\
 & & & & & & & \\
$70\um$ & MJy/sr & 0--60 & 
	1426 & $0.43\pm0.02$ & $2.69\pm0.13$ & $0.39\pm0.01$ & 1.6 \\
 & & 0--35 & 502 & $0.41\pm0.03$ & $2.57\pm0.18$ & $0.37\pm0.02$ & 1.8 \\
 & & 35--60 & 924 & $0.43\pm0.02$ & $2.69\pm0.13$ & $0.39\pm0.01$ & 1.3 \\
 & & & & & & & \\
$160\um$ & MJy/sr & 0--60 & 1516 & 
	$1.36\pm0.01$ & $23.4\pm0.6$ & $0.26\pm0.01$ & 1.8 \\
 & & 0--35 & 535 & $1.30\pm0.01$ & $20.0\pm0.5$ & $0.26\pm0.01$ & 1.5 \\
 & & 35--60 & 981 & $1.40\pm0.01$ & $25.1\pm0.6$ & $0.26\pm0.01$ & 1.4 \\
 & & & & & & & \\
\Av\ & mag. & 0--60 & 1406 & $-0.04\pm0.02$ & $0.91\pm0.04$ & $0.20\pm0.01$ & 6.6 \\
 & & 0--35 & 495 & $-0.16\pm0.02$ & $0.69\pm0.03$ & $0.25\pm0.01$ & 1.9 \\
 & & 35--60 & 910 & $0.01\pm0.01$ & $1.02\pm0.03$ & $0.18\pm0.01$ & 4.4 \\
\hline\noalign\\  
\end{tabular}

\medskip
$Y$ is the fraction of sightlines in each bin divided by the logarithmic bin-
width $\dif(\log_{10}X)$. $\chi^2$ is the reduced chi-squared goodness of fit 
parameter, with the error in each bin $\delta_i$ estimated as $\delta_i=
\sqrt{N_i}$ and for (number of bins$-2$) degrees of freedom.
\label{table:M31:dust}
\end{table*}

We calculated the PDFs of the dust emission from M31 at $24\,\um$,
$70\,\um$ and $160\,\um$ using the Spitzer MIPS maps of
\citet{Gordon:2006} smoothed to a resolution of 45\,\arcsec\, by
\citet{Tabatabaei:2010}. The lognormal fits are given in
Table~\ref{table:M31:dust}. At $24\,\um$ and $70\,\um$ the fits for the 3
regions closely agree, as is also the case for the PDFs of \NHH\ and
\Ha\ (see Table~\ref{table:M31:surface}). The dispersion of $0.34\pm
0.01$ of the $24\,\um$ PDFs is the same as that of \Ha\ and the
dispersion of $70\,\um$ PDFs agrees with that of \NHH. As both the
ionized gas and the warm dust are mainly heated by young OB stars, the
agreement in dispersion suggests that the density structure of the
ionized gas is similar to that of the warm dust particles. This is
consistent with the good pixel-to-pixel correlation between these dust
emissions and that of ionized gas reported by \citet{Tabatabaei:2010}.

Although at $160\,\um$ the dispersion of the PDFs is also the same for
each region, the value of $X_0$ for the 'ring' is about 25 per cent higher
than for the inner region. The dispersion of $0.27\pm 0.01$ is close
to that of the \HI\ PDFs (see Table~\ref{table:M31:surface}). Since
the dust emitting at $160\,\um$ is mainly heated by the smooth ISRF
\citep{Xu:1996}, this suggests that the density structures of the cool
dust and the \HI\ gas are similar. Indeed, \citet{Tabatabaei:2010}
found that the wavelet correlation on scales below about $1\,\kpc$
between $160\,\um$ and \HI\ is much better than between the other
wavelengths and \HI.

We also calculated the PDFs of the dust extinction \Av\ through M31
using the map of optical depth at the wavelength \Ha, $\tau_{\Ha}$,
derived by \citet{Tabatabaei:2010}, and $\Av = 1.234\,\tau_{\Ha}$. The
parameters of the lognormal fits are given at the end of
Table~\ref{table:M31:dust}. The dispersions in the \Av\ PDFs for the
subregions closely agree with those of the column densities of \HI\
(see Table~\ref{table:M31:surface}), the dominant gas phase in M31.
This indicates that in spite of the radial increase in the atomic gas-
to-dust ratio the density structure in the atomic gas and the cool 
dust are very similar, in agreement with the nearly linear correlation 
between extinction and \NHI\ obtained by \citet{Tabatabaei:2010}.

%________________________________SECTION   5.6____________________________

\subsection{Comparison with the Milky Way}
\label{subsec:MW}

The \Ha\ PDFs for M31 and M51 may be compared to those obtained
for the MW by \citet{Hill:2008} from the Wisconsin \Ha-Mapper
survey \citep{Haffner:2003}. For the PDF of the emission measure
perpendicular to the Galactic plane, EM~$\sin{|b|}$, of the diffuse ionized 
gas (DIG) at $|b|>10\degree$ they found a dispersion of $0.190\pm 0.001$, 
which is nearly half the value for the DIG in M51 \citep{Seon:2009} and
for M31 (see Table~\ref{table:M31:surface}). However, in order to derive 
the PDF of DIG only, \citet{Hill:2008} removed all sight-lines towards 
classical \HII\ regions from their sample, whereas in our PDFs for M31 and 
M51 these regions are included. The MW PDF including classical \HII\ regions 
has a much larger dispersion than 0.19 and shows a high-density excess like 
the \Ha\ PDFs for M31 and M51. High-density tails are expected 
\citep{Ostriker:2001} if the temperature in the medium decreases with 
increasing density, as is the case in the ionized medium in the MW 
\citep{Madsen:2006}. 

\citet{Hill:2008} showed that the dispersion of the PDF of EM~$\sin{|b|}$
decreases with increasing latitude. \citet{Seon:2011} found the same effect
for the diffuse background emission from the MW in FUV, which mainly consists
of starlight scattered by interstellar dust. Because the LOS decreases with
increasing $|b|$, we would expect the opposite behaviour (see 
Sect.~\ref{subsec:LOS}). However, the decrease in $\sigma$ is consistent with 
the observation of \citet{Berkhuijsen:2008b} that the spread in $\avg{n_e}$ 
decreases with increasing distance to the plane $|z|$ (also visible in 
Fig. 5 of \citet{Gaensler:2008}), which is accompanied by an increase in the 
size of ionized clouds. \citet{ Savage:2009} arrived at a similar conclusion 
based on \OVI-line observations. We show below that the decrease in dispersion 
with increasing $|b|$ is due to an increase in the volume filling factors of 
the ionized gas and the dust grains, leading to an increase in the effective 
LOS.

\citet{Hill:2008} and \citet{Seon:2011} derived the dispersion for the 
latitude intervals $10\degree<|b|<30\degree$, $30\degree<|b|<60\degree$ and 
$60\degree<|b|<90\degree$. From the dispersion and the scaling explained in
Sect.~\ref{subsec:fv}, we calculated $\Leff$\, for each case, which increases 
when $\sigma$ decreases. Assuming a maximum LOS equal
to the scale height of the DIG of $L=1\,\kpc$ \citep{Berkhuijsen:2008b,
Schnitzeler:2012}, we obtain the volume filling factor $f_v = \Leff/L$.
Table~\ref{table:MW:leff} shows that $f_v$ of both the DIG and the scattering
dust increases with increasing $|b|$. The largest part of the DIG and dust 
layers is sampled by LOS in the interval $10\degree<|b|<30\degree$, whereas
at the highest latitudes mainly DIG and dust at high $|z|$ near the Sun are
observed. Therefore, the value of $f_v=0.26\pm0.05$ for the DIG at 
$10\degree<|b|<30\degree$ best represents the mean filling factor of the 
DIG seen towards $L=1\,\kpc$ perpendicular to the Galactic plane. This value 
agrees well with the value of $f_v=0.22\pm0.05$ obtained by 
\citet{Berkhuijsen:2006} from emission measures and dispersion measures 
towards 157 pulsars at $|b|>10\degree$. These autors found an increase of the 
\emph{local} filling factor $f_v(z)$ with increasing $|z|$, which was 
confirmed by \citet{Gaensler:2008} using DM and EM towards 51 pulsars with 
known distance. An increase of $f_v(z)$ with $|z|$ is also indicated by 
the increase of the mean $f_v$ towards higher latitudes in 
Table~\ref{table:MW:leff}.

The dust filling factors for the two lower intervals of $|b|$ are about half
of those of the DIG, but at high $|b|$ the values of $f_v$ are the same.
Near the plane the clumpiness of the dust seems larger than that of 
the DIG. On the other hand, the values of $f_v$ will be too low if the
scale height of the diffuse dust is smaller than 1\,\kpc.

%________________TABLE 7_______ \label{table:MW:leff}_________________________

\begin{table}
\caption{Effective LOS and volume filling factors towards $|z|=1\kpc$ in the MW.}
\begin{tabular}{llrrr}
\hline
 Data & $|b|$ & $\sigma$ & $\Leff$ & $f_v$  \\
      & (degree) & & [pc] &  \\
\hline
EM~$\sin{|b|}$ & 10-30 & 0.19 & $260\pm50$ & $0.26\pm0.05$ \\
               & 30-60 & 0.16 & $390\pm70$ & $0.39\pm0.07$ \\
               & 60-90 & 0.14 & $490\pm90$ & $0.49\pm0.09$ \\
 & & & & \\
$I_{FUV}$~$\sin{|b|}$  & 10-30 & 0.28 & $120\pm20$ & $0.12\pm0.02$ \\
                       & 30-60 & 0.23 & $180\pm30$ & $0.18\pm0.03$ \\
                       & 60-90 & 0.14 & $490\pm90$ & $0.49\pm0.09$ \\
\hline\noalign\\
\end{tabular}
\medskip

Only random errors due to the error in $\sigma$ of $\avg{n_e}$ are given (see
\ref{table:los}). Errors in $\sigma$ of EM~$\sin{|b|}$ are negligible. Errors 
in $\sigma$ of $I_{FUV}$~$\sin{|b|}$ are not available.
\label{table:MW:leff}
\end{table}

%___________________________SECTION  6____________________________________

\section{Summary}
\label{sec:summary}

We have presented probability distribution functions (PDFs) of the surface 
densities of neutral and ionized gas at radii $R\lesssim 3\arcmin$ in M51 
and in the radial ranges $0\arcmin<R<35\arcmin$, $35\arcmin<R<60\arcmin$ 
and $0\arcmin<R<60\arcmin$ in M31 (Section~\ref{subsec:surface}, Figs.~\ref
{fig:M31} \& \ref{fig:M51}). For M31 we also presented the PDFs of the 
average volume densities along the LOS of \HI, \HH\ and total gas (Fig.~
\ref{fig:M31:ring}) and derived the PDFs of the dust emission and extinction 
(Table~\ref{table:M31:dust}). The main conclusions from Sect.~\ref{sec:PDFs}, 
Sect.~\ref{sec:models} and  the discussion in Sect.~\ref{sec:discussion}
may be summarized as follows.

1. All PDFs are close to lognormal, but their dispersions differ
(Tables~\ref{table:M31:surface}--\ref{table:M31:dust}). We have investigated 
which factors determine the dispersion. 

2. The dispersion of the lognormal PDFs of column and average volume densities
depend on the pathlength $L$, with $\sigma\propto L^{-0.5}$, and telescope 
resolution $D$, with $\sigma\propto D^{-1}$. The PDFs of column and
average volume density are also affected by the filling factor of the gas; 
if an effective LOS is defined as $\Leff=f_{v} L$ then 
$\sigma\propto \Leff^{-0.5}$.

3. The length of the line of sight through the medium, $L$, is the dominant 
factor shaping the PDF. Short LOS through the nearly face-on galaxy M51 cause 
wider PDFs than long LOS through the nearly edge-on galaxy M31, and for both 
galaxies the \HH\ PDFs are wider than the \HI\ PDFs because the scale height 
of molecular gas is only half that of atomic gas. The contribution of the 
higher Mach number in the \HH\ to the widening of the observed PDFs needs 
further study.

4. The dispersions of the PDFs of \HI\ and \HH\ column densities in M31 are 
independent of the beamwidth for angular resolutions of 23\,arcsec to 
180\,arcsec (Fig.~\ref{fig:M31:disp}) corresponding to linear resolutions in 
the plane of M31 of $90\,\pc \times 400\,\pc\,$ to $700\,\pc \times 3150\,\pc
\,$ along major x minor axis. This suggests that the maximum cloud size 
$\turb \lesssim 95\,\pc$.

5. The dispersions of the volume-density PDFs of \HI\ and \HH\ for M31 
(Fig.~\ref{fig:M31:ring}), of diffuse \HI\, and diffuse ionized gas for 
the solar neighbourhood \citep{Berkhuijsen:2008, Berkhuijsen:2012}, and of 
diffuse ionized gas for M51 vary with $L$ as 
$\sigma \propto L^{-0.76\pm0.06}$ (Fig.~\ref{fig:los}), indicating that 
the density structures are very similar in M31, M51 and the Milky Way.

6. The exponent of the power law $\sigma$--$L$ relation is steeper than 
the $-0.5$ expected from analytic calculations \citep{Fischera:2004} and the 
models in Sect.~\ref{sec:models}. We show that the difference can be explained 
by an increase in the volume filling factor between dense and less dense 
regions causing an increase in the effective LOS, $\Leff=f_{v} L$. An increase 
in $f_v$ also explains the decrease in $\sigma$ with increasing latitude of 
the PDFs observed for the DIG and the scattering dust in the Milky Way 
\citep{Hill:2008, Seon:2011}. 

7. The dispersions of the \Av\ PDFs for M31 closely agree with those of the 
column densities of \HI, the dominant gas phase in M31, suggesting that the 
density structures in the cool dust and in \HI\ are similar.

8. The lognormal PDFs we obtained for M31 and M51
do not, taken in isolation, show that compressible turbulence is shaping the 
gas density distribution. However, they are compatible with this interpretation:
both in terms of their shape and in how their properties vary with beamwidth and 
LOS. By combining the PDF parameters with independent measures of, for example, 
the Mach number, filling factor and plasma beta, a deeper understanding of 
turbulence in a multi-phase interstellar medium should become possible.

%__________________________________________________________________________
\section{Acknowledgements}
AF thanks the Leverhulme Trust (RPG-097) and the STFC (ST/L005549/1) for 
financial support. We thank Dr. Sui Ann Mao for useful comments on the 
manuscript. We also thank Dr. Christoph Federrath for interesting comments
and suggestions that led to improvements in the manuscript.

%__________________________________________________________________________

\bsp

\label{lastpage}

\end{document}